\documentclass[twocolumn,showpacs,superscriptaddress,aps,pre,10pt]{revtex4-1}

\usepackage{graphicx}
\usepackage{amssymb,amsmath,dcolumn}
\usepackage{bm}
\usepackage{color}

\usepackage[bookmarks=true,colorlinks,linkcolor=blue,anchorcolor=blue,citecolor=blue,unicode]{hyperref} %

\usepackage{bookmark}

\begin{document}

\preprint{Fractals/HFTN}

\title{Highway freight transportation diversity of cities based on radiation models}

\author{Li Wang}
 \affiliation{School of Business, East China University of Science
   and Technology, Shanghai 200237, China}

\author{Jun-Chao Ma}
 \affiliation{School of Business, East China University of Science
   and Technology, Shanghai 200237, China}
 \affiliation{Zhicang Technologies, Beijing 100016, China}  
   
\author{Zhi-Qiang Jiang}
 \affiliation{School of Business, East China University of Science
   and Technology, Shanghai 200237, China}
 \affiliation{Research Center for Econophysics, East China University of
   Science and Technology, Shanghai 200237, China}

\author{Wanfeng Yan}
 \email{wanfeng.yan@gmail.com}
 \affiliation{Zhicang Technologies, Beijing 100016, China}  
 \affiliation{Research Center for Econophysics, East China University of
   Science and Technology, Shanghai 200237, China}

\author{Wei-Xing Zhou}
 \email{wxzhou@ecust.edu.cn}
 \affiliation{School of Business, East China University of Science
   and Technology, Shanghai 200237, China}
 \affiliation{Research Center for Econophysics, East China University of
   Science and Technology, Shanghai 200237, China}
 \affiliation{Department of Mathematics, East China University of
   Science and Technology, Shanghai 200237, China}

\date{\today}

\begin{abstract}

Using a unique data set containing about 15.06 million truck transportation records in five months, we investigate the highway freight transportation diversity of 338 Chinese cities based on the truck transportation probability $p_{ij}$ from one city to the other. The transportation probabilities are calculated from the radiation model based on the geographic distance and its cost-based version based on the driving distance as the proxy of cost. For each model, we consider both the population and the gross domestic product, and find quantitatively very similar results. We find that the transportation probabilities have nice power-law tails with the tail exponents close to 0.5 for all the models. The two transportation probabilities in each model fall around the diagonal $p_{ij}=p_{ji}$ but are often not the same. In addition, the corresponding transportation probabilities calculated from the raw radiation model and the cost-based radiation model also fluctuate around the diagonal $p_{ij}^{\rm{geo}}=p_{ij}^{\rm{cost}}$. We calculate four sets of highway truck transportation diversity according to the four sets of transportation probabilities that are found to be close to each other for each city pair. Further, it is found that the population, the gross domestic product, the in-flux, and the out-flux scale as power laws with respect to the transportation diversity in the raw and cost-based radiation models. It implies that a more developed city usually has higher diversity in highway truck transportation, which reflects the fact that a more developed city usually has a more diverse economic structure. 


\end{abstract}


\maketitle


\section{Introduction}
\label{S1:Intro}

Aviation, railway, highway and shipping are four main transportation methods in modern societies. Unlike other three ones, information about highway transportation is less publicly available. In mainland China, the highway system has experienced a very rapid development since the Reform and Opening-up of China, forming a rapidly expanding multiplex network which contains national highways, provincial highways, county highways and countryside highways \cite{Wang-Ma-Jiang-Yan-Zhou-2019-EPJDataSci}. China has the longest expressway network in the world, which includes about 0.143 million kilometers expressways.

In the past decades, the gravity law is the most adopted in understanding transportation networks and predicting transportation fluxes \cite{Jung-Wang-Stanley-2008-EPL,Masucci-Serras-Johansson-Batty-2013-PhysRevE,Lenormand-Bassolas-Ramasco-2016-JTranspGeogr,Barbosa-Barthelemy-Ghoshal-James-Lenormand-Louail-Menezes-Ramasco-Simini-Tomasini-2018-PhysRep,Piovani-Arcaute-Uchoa-Wilson-Batty-2018-RSocOpenSci}, which reads
\begin{equation}
  W_{ij} \sim \frac{M_i^{\alpha}M_j^{\beta}}{d_{ij}^{\gamma}},
\end{equation}
where $W_{ij}$ is the flow between locations $i$ and $j$, $M_i$ (or $M_j$) is usually the population or gross domestic product (GDP) of location $i$ (or $j$), $d_{ij}$ is the distance between $i$ and $j$, and $\alpha$, $\beta$ and $\gamma$ are the model parameters.
Very relevantly, the gravity law has been investigated and confirmed in the Korean highway network between 30 largest cities \cite{Jung-Wang-Stanley-2008-EPL}, the express bus flow in Korea consisting of 74 cities and 170 bus routes with 6692 operating buses per day \cite{Kwon-Jung-2012-PA}, and the urban bus networks of Korean cities \cite{Hong-Jung-2016-PA}, and the highway freight transportation networks of 338 Chinese cities \cite{Wang-Ma-Jiang-Yan-Zhou-2019-EPJDataSci}.
 
However, the gravity model has several limitations, especially the requirement of previous traffic data to fit the parameters \cite{Simini-Gonzalez-Maritan-Barabasi-2012-Nature}. To overcome those limitations, the radiation model has been proposed \cite{Simini-Gonzalez-Maritan-Barabasi-2012-Nature}, in which the predicted flux $\tilde{F}_{ij}$ from city $i$ to city $j$ is obtained as follows
\begin{equation}
  {\tilde{F}_{ij}} = F_i^{\rm{out}}\frac{M_iM_j}{(M_i+S_{ij})(M_i+M_j+S_{ij})},
  \label{Eq:Radiation:Fij}
\end{equation}
where $S_{ij}$ is the total ``mass'' (population or GDP) in the circle of radius $d_{ij}$ centered at $i$ but excluding the source and destination population, and $F_i^{\rm{out}}$ is total out-flux departing from city $i$
\begin{equation}
  F_i^{\rm{out}} = \sum_{j\neq{i}} F_{ij},
\end{equation}
where $F_{ij}$ is the real flux from $i$ to $j$. Obviously, the data of $F_i^{\rm{out}}$ are much easier to collect than $F_{ij}$. 

In the raw radiation model, $d_{ij}$ is the geographic distance between $i$ and $j$. The cost-based radiation model has been soon proposed based on the intuition that an individual will choose the site that has the lowest travel cost on the network, where the travel cost can be measured by the path length or travel time from $i$ to $j$ \cite{Ren-ErcseyRavasz-Wang-Gonzalez-Toroczkai-2014-NatCommun}. In this work, $d_{ij}$ is measure by the path length or driving distance from $i$ to $j$. Later, to better estimate the fluxes at different spatial scales, a scaling parameter is introduced into the radiation model \cite{Yang-Herrera-Eagle-Gonzalez-2014-SciRep}. By combining memory effect and population-induced competition, a general model has been developed to enable accurate prediction of human mobility based on population distribution only, which also has a parameter qualifying the memory effect \cite{Yan-Wang-Gao-Lai-2017-NC}.


Although the radiation model has been adopted in the study of trip distributions \cite{Lenormand-Huet-Gargiulo-Deffuant-2012-PLoSOne,Gargiulo-Lenormand-Huet-Espinosa-2012-JASSS,Lenormand-Huet-Gargiulo-2014-JTranspLandUse,Lenormand-Bassolas-Ramasco-2016-JTranspGeogr,Xia-Cheng-Chen-Wei-Zong-Li-2018-JTranspGeogr}, applications to freight transportation are rare. In this work, using a unique data set about the freight highway transportation by trucks between 338 cities in mainland China, we investigate the transportation probability $p_{ij}$ between two cities $i$ and $j$ and the transportation diversity of a city defined by $p_{ij}$. Although most studies dealt with undirected transportation networks \cite{Serrano-Boguna-2003-PRE,Garlaschelli-Loffredo-2004-PRL,Wang-Ma-Jiang-Yan-Zhou-2019-EPJDataSci}, radiation models enable us to consider directed transportation networks due to the availability of data \cite{Wang-Ma-Jiang-Yan-Zhou-2019-EPJDataSci}. The raw radiation model and the cost-based radiation model are adopted because they are parameter free. Our analysis shows that the population, the gross domestic product, the in-flux, and the out-flux scale as power laws with respect to the transportation diversity in the raw and cost-based radiation models, which implies that a more developed city usually has higher diversity in highway truck transportation. This finding reflects the fact that a more developed city usually has a more diverse economic structure.

The remainder of this work is organized as follows. Section~\ref{S1:Data} describes the data sets we analyze. Section~\ref{S1:Radiation:pij} studies the basic properties of transportation probability. Section \ref{S1:Diversity} deals with the transportation diversity of cities and their relationship with population and GDP. We discuss and summarize in Section \ref{S1:Summary}.

\section{Data sets}
\label{S1:Data}

The data set we analyze was provided by a leading truck logistics company in China, which records the highway truck freight transportation between 338 cities in mainland China over the period from 1 January 2019 to 31 May 2019 \cite{Wang-Ma-Jiang-Yan-Zhou-2019-EPJDataSci}. There are about 15.06 million truck freight transportation records in total, each entry containing the origin and destination cities and the starting date of the transportation. We can construct the flux matrix ${\mathbf{F}} = \left[F_{ij}\right]_{338\times338}$, where $F_{ij}$ stands for the number of trucks with freights driven from city $i$ to city $j$. Unloaded trucks are not counted in. Because radiation models do not consider intra-city transportation, we set that
\begin{equation}
 F_{ii}=0.
\end{equation}
It is obvious that $F_{ij}$ is not necessary to be equal to $F_{ji}$ for $i\neq{j}$.

The GDP and population data for the 338 Chinese cities in 2017 were retrieved online from the Complete Collection of World Population (http://www.chamiji.com), which are publicly available except for a few cities. We supplemented the missing data by searching Baidu Encyclopedias (https://baike.baidu.com).

The geographic distance $d_{ij}^{\rm{geo}}$ is the shortest surface distance between two cities located by the longitude and latitude, which is the length of the great circle arc connecting two points on the surface of the earth. The longitude and latitude of each city can be easily obtained online for free. The data set of the driving distances  $d_{ij}^{\rm{cost}}$ between pairs of cities was provided by the same truck logistics company, which were collected by their truck drivers. The driving distance between two cities are usually ``optimized'' by the truck drivers because they always have the motivation to find a path connecting the two cities with the least cost (time and money). Such an optimization is achieved either by their own experience or by information from buddy truck drivers they trust. It is obvious that
\begin{equation}
  d_{ij}^{\rm{geo}} < d_{ij}^{\rm{cost}}
  \label{Eq:dgeo:lt:dcost}
\end{equation}
for all pairs of cities. The difference between these two distances increases when the two cities are farther away to each other. By definition, the geographic distance matrix is symmetric, that is,
\begin{equation}
  d_{ij}^{\rm{geo}}=d_{ji}^{\rm{geo}}.
\end{equation}
In contrast, the driving distance matrix is asymmetric, i.e., 
\begin{equation}
 d_{ij}^{\rm{cost}} \neq d_{ji}^{\rm{cost}},
\end{equation}
which is mainly due to the fact that, besides highways, there are often local roads that a truck driver has to take from one city to the other.

\begin{figure*}[!t]
  \centering
  \includegraphics[width=0.46\linewidth]{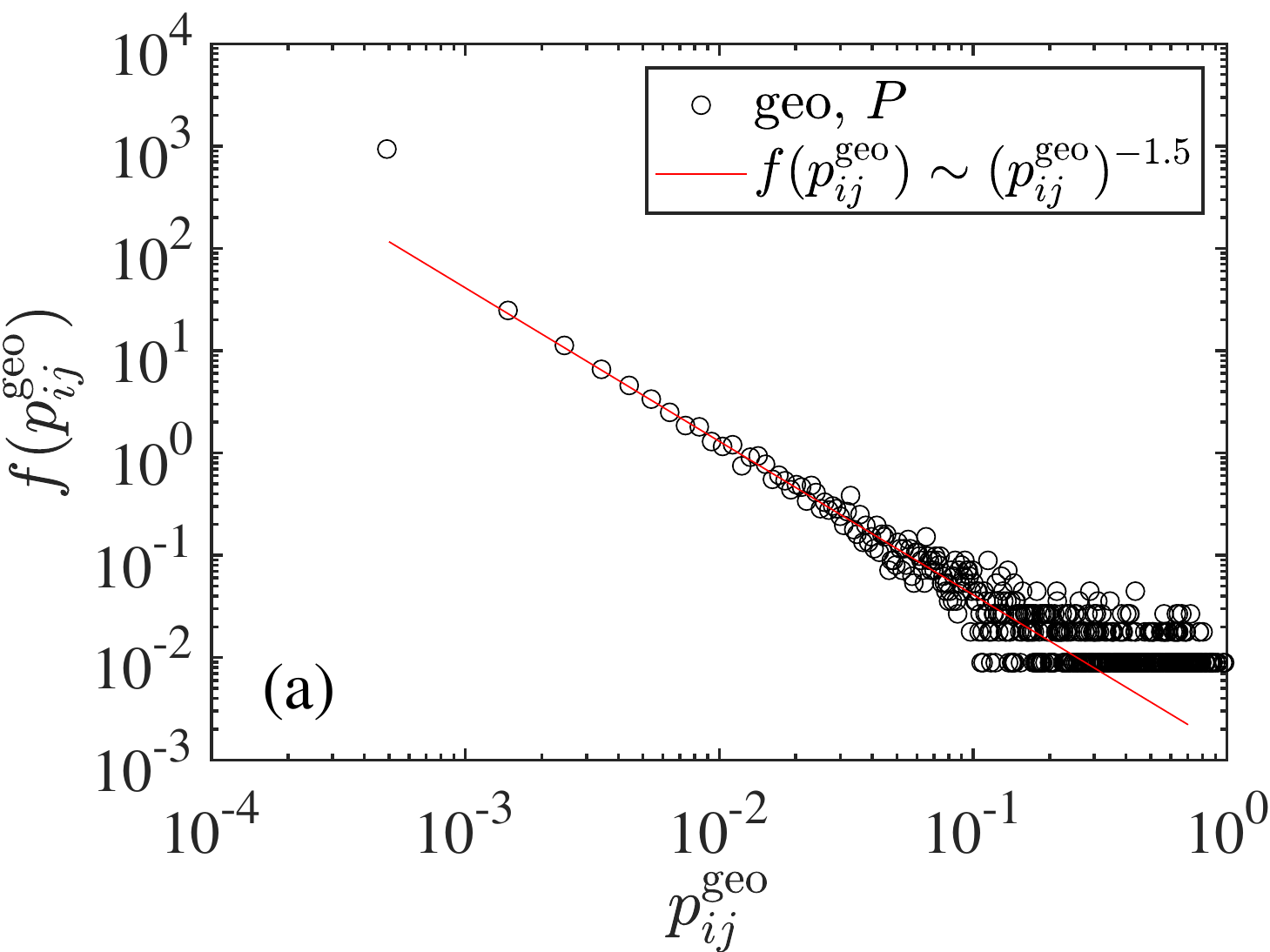}   
  \includegraphics[width=0.46\linewidth]{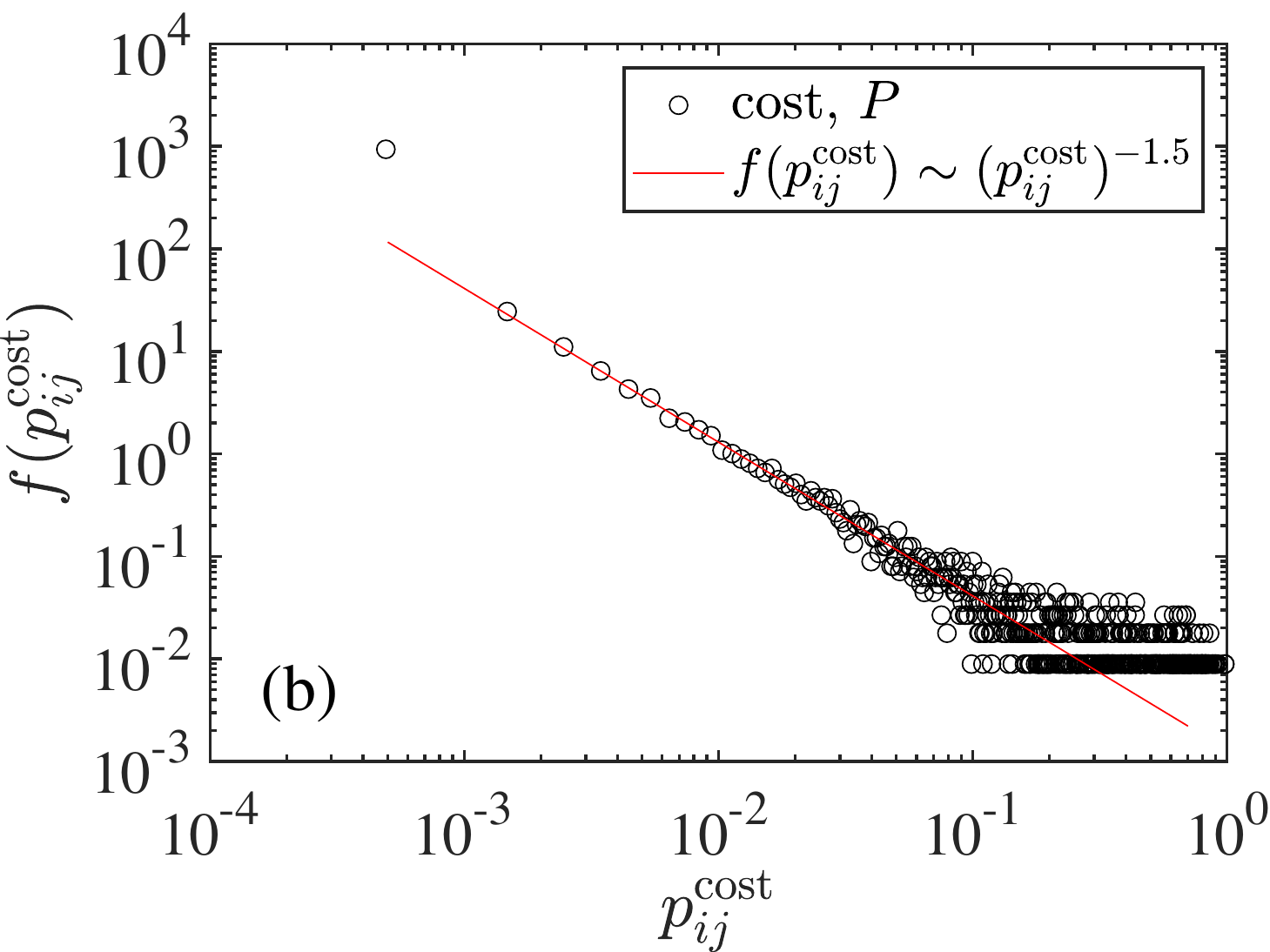}\\    
  \includegraphics[width=0.46\linewidth]{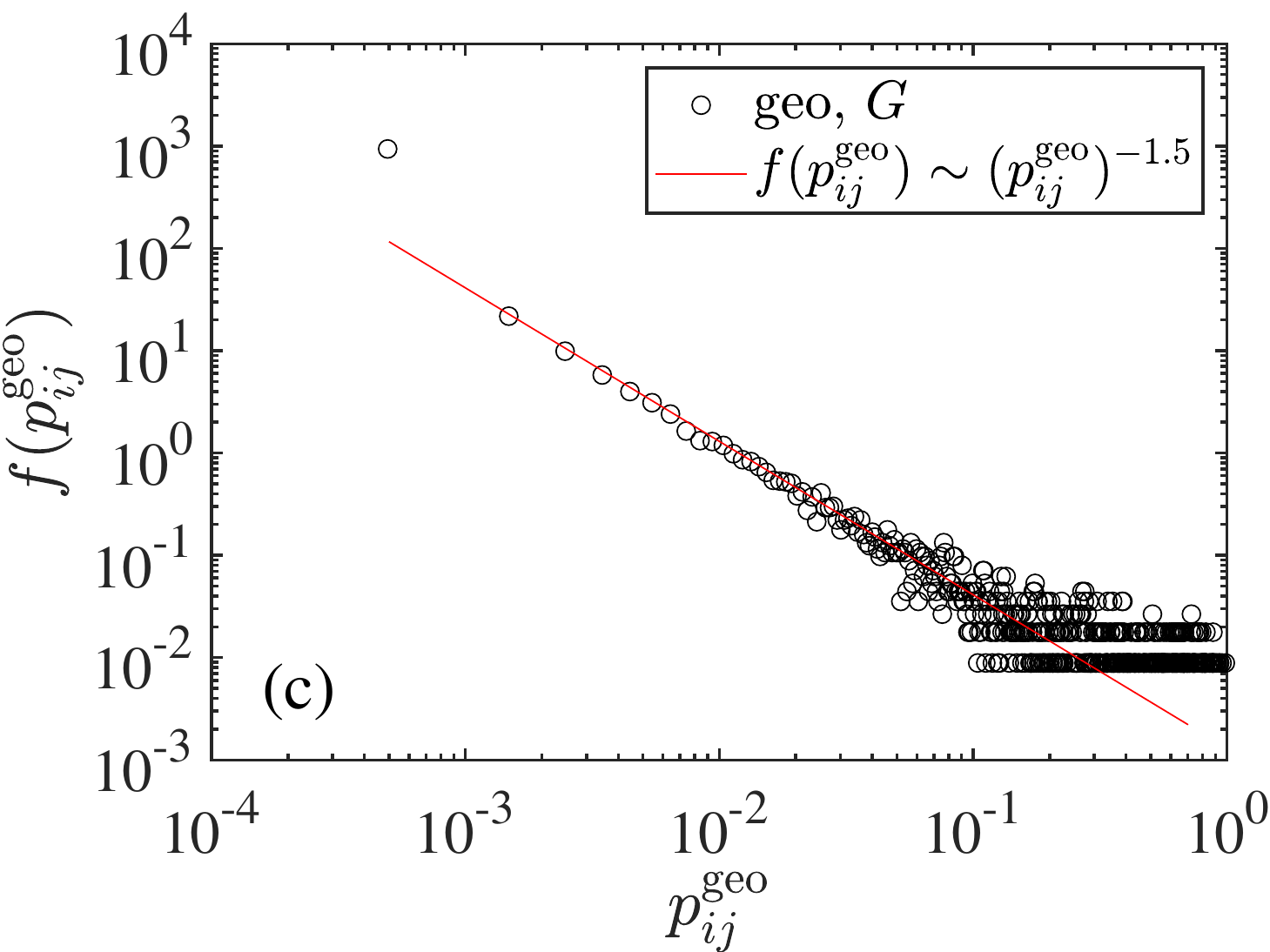} 
  \includegraphics[width=0.46\linewidth]{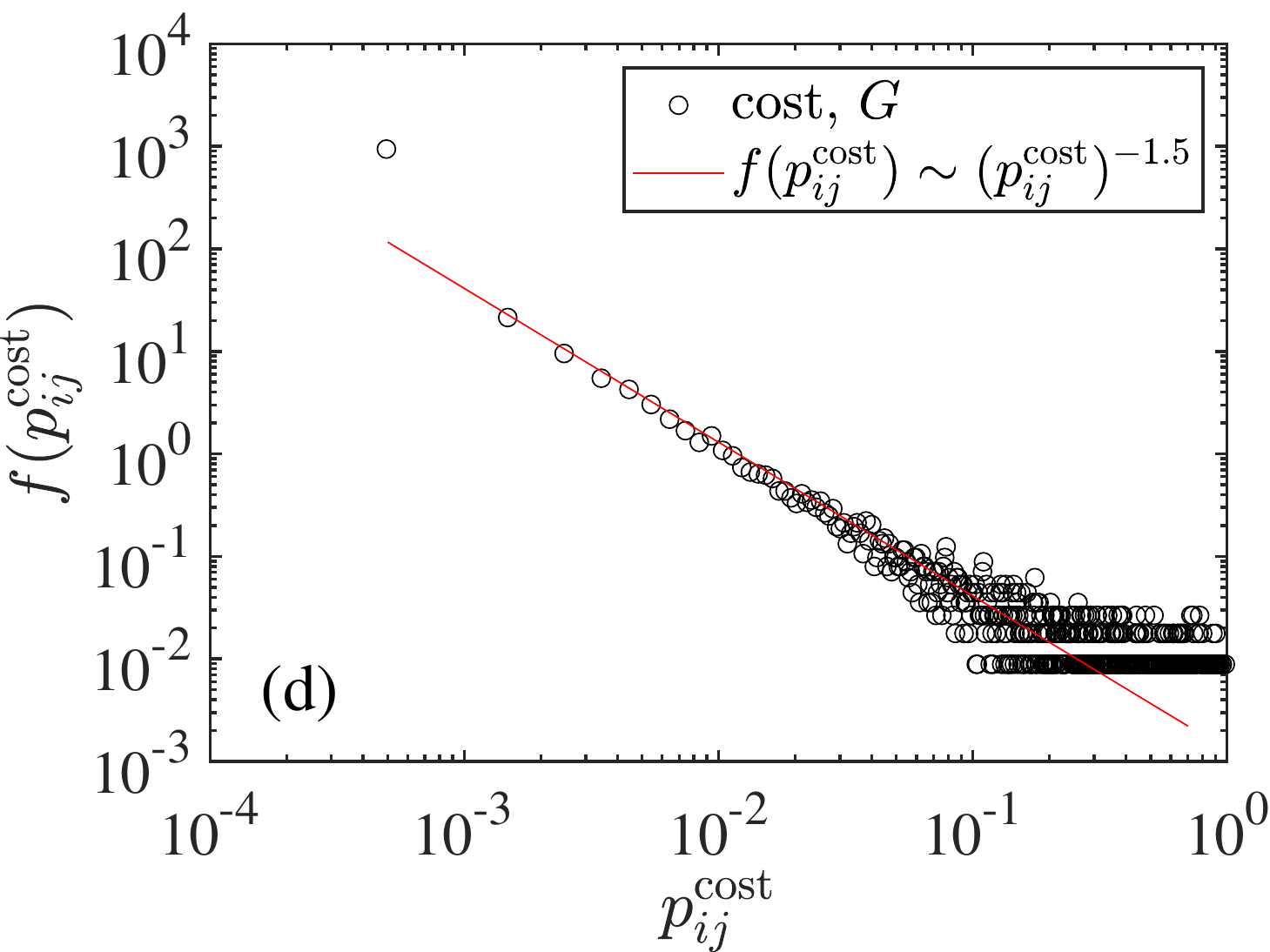}       
  \caption{Power-law tailed distribution of the transportation probability between two cities. The solid lines are power laws with the same exponent of -1.5. (a) Population $P$ is used in the raw radiation model with the geographic distance. (b) Population $P$ is used in the cost-based radiation model with the driving distance. (c) Gross domestic product $G$ is used in the raw radiation model with the geographic distance. (d) Gross domestic product $G$ is used in the cost-based radiation model with the driving distance. }
  \label{Fig:PDF:pij}
\end{figure*}

\section{Transportation probability}
\label{S1:Radiation:pij}

\subsection{Formulae}

According to the radiation models (\ref{Eq:Radiation:Fij}) we adopt, the transportation probability $p_{ij}$ from city $i$ to city $j$ is
\begin{equation}
  p_{ij} = \frac{M_iM_j}{(M_i+S_{ij})(M_i+M_j+S_{ij})}.
  \label{Eq:Radiation:pij}
\end{equation}
When we choose population $P$ for $M$, the transportation probability becomes
\begin{equation}
  p_{ij} = \frac{P_iP_j}{(P_i+S_{ij})(P_i+P_j+S_{ij})},
  \label{Eq:Radiation:pij:P}
\end{equation}
where $S_{ij}$ is the total population in the circle of radius $d_{ij}$ centered at $i$ but excluding the source and destination population.
Alternatively, when we use GDP as the proxy, we have
\begin{equation}
  p_{ij} = \frac{G_iG_j}{(G_i+S_{ij})(G_i+G_j+S_{ij})},
  \label{Eq:Radiation:pij:G}
\end{equation}
where $S_{ij}$ is the total GDP in the circle of radius $d_{ij}$ centered at $i$ but excluding the source and destination population.

The transportation probabilities $p_{ij}$ of the raw radiation model using geographic distance and the cost-based radiation model using driving distance are calculated with respect to population $P$ in Eq.~(\ref{Eq:Radiation:pij:P}) and gross domestic product $G$ in Eq.~(\ref{Eq:Radiation:pij:G}).

\subsection{Power-law distribution of $p_{ij}$}

Figure~\ref{Fig:PDF:pij} illustrates the four empirical distributions of the transportation probability $p_{ij}$ between two cities for the two radiation models with $M=P$ and $M=G$ respectively. We observe a nice power-law tail in each case and the exponents are the same for the four cases:
\begin{equation}
  f(p_{ij}) \sim p_{ij}^{-\alpha-1},
\end{equation}
where the tail exponents $\alpha\approx 0.5$ and the intercepts are almost the same. The power-law relationship holds over three orders of magnitude. The smallest transportation probabilities deviate from the power-law distributions with higher probability density. Theoretically, we know that two cities with longer distance usually have a smaller transportation probability. Indeed, it we plot $p_{ij}$ with respect to $d_{ij}$, we find that the points fluctuate around a power-law scaling with an exponent of $-4$:
\begin{equation}
  p_{ij} \sim d_{ij}^{-4},
\end{equation}
which corresponds to the case of uniform population (or GDP) density \cite{Simini-Gonzalez-Maritan-Barabasi-2012-Nature}. The standard deviation of the data points from this reference power law quantifies the strength of heterogeneity of the spatial distribution of population and GDP in mainland China.

\begin{figure*}[!t]
  \centering
  \includegraphics[width=0.46\linewidth]{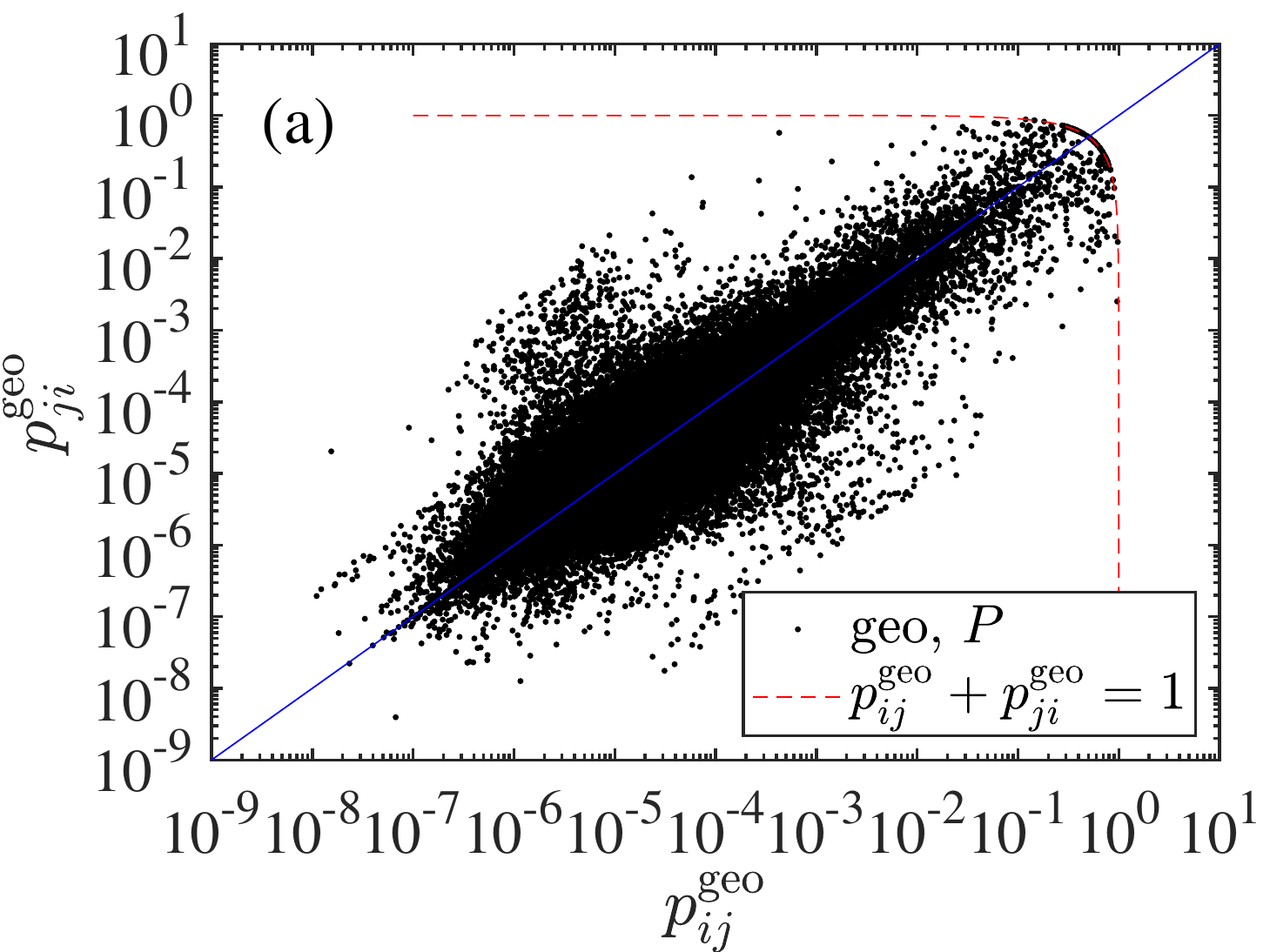}  
  \includegraphics[width=0.46\linewidth]{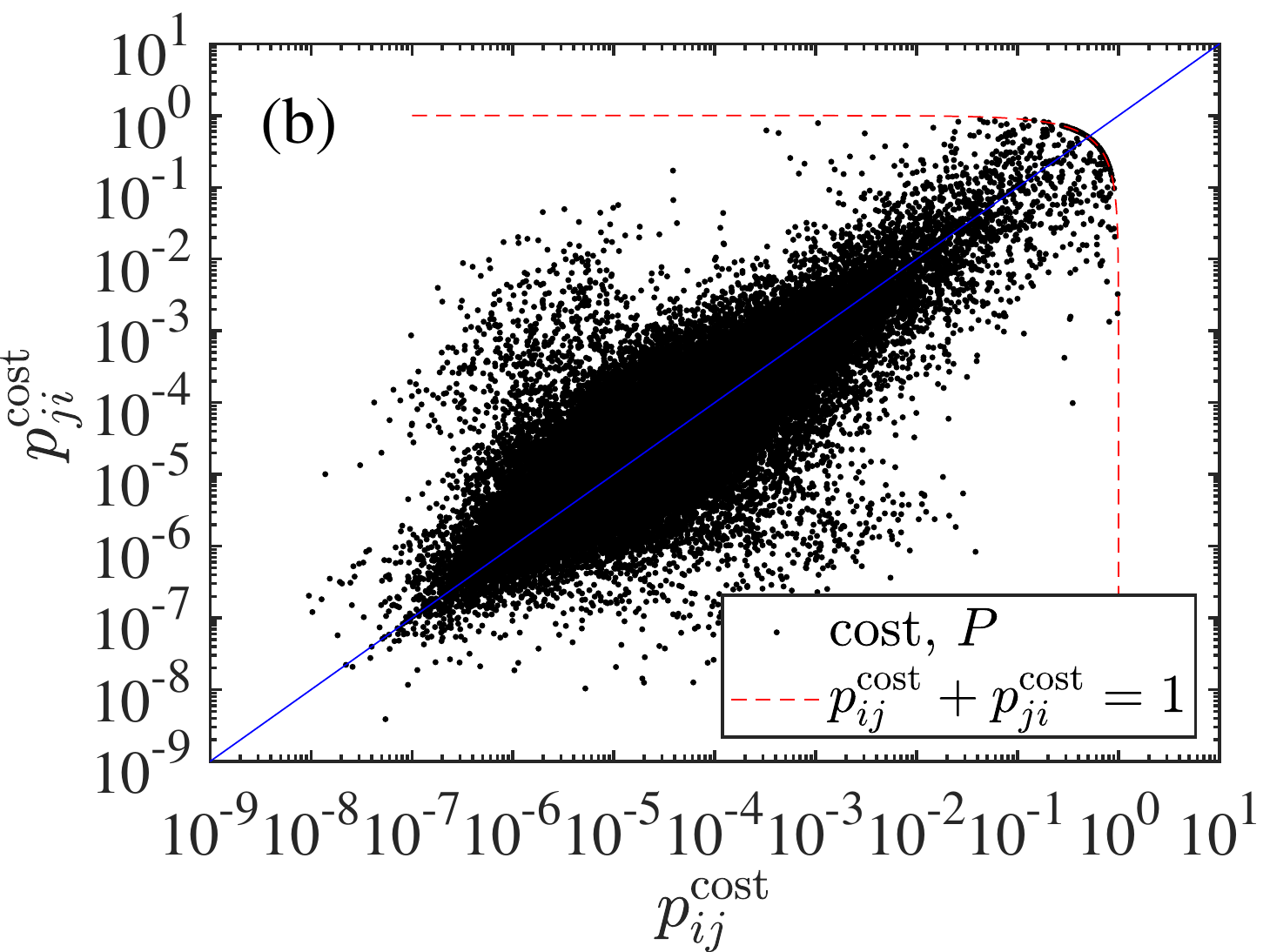}   \\
  \includegraphics[width=0.46\linewidth]{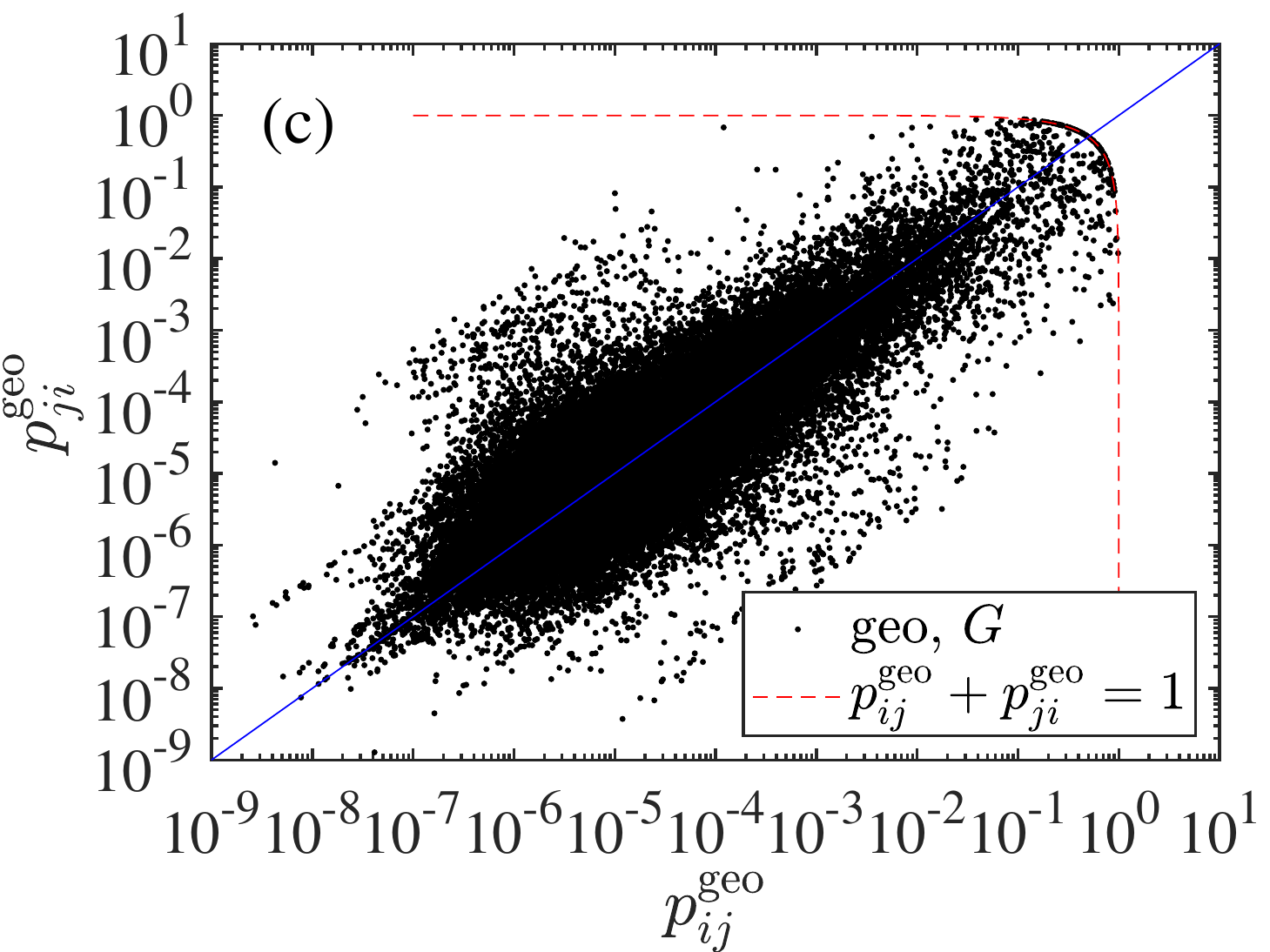} 
  \includegraphics[width=0.46\linewidth]{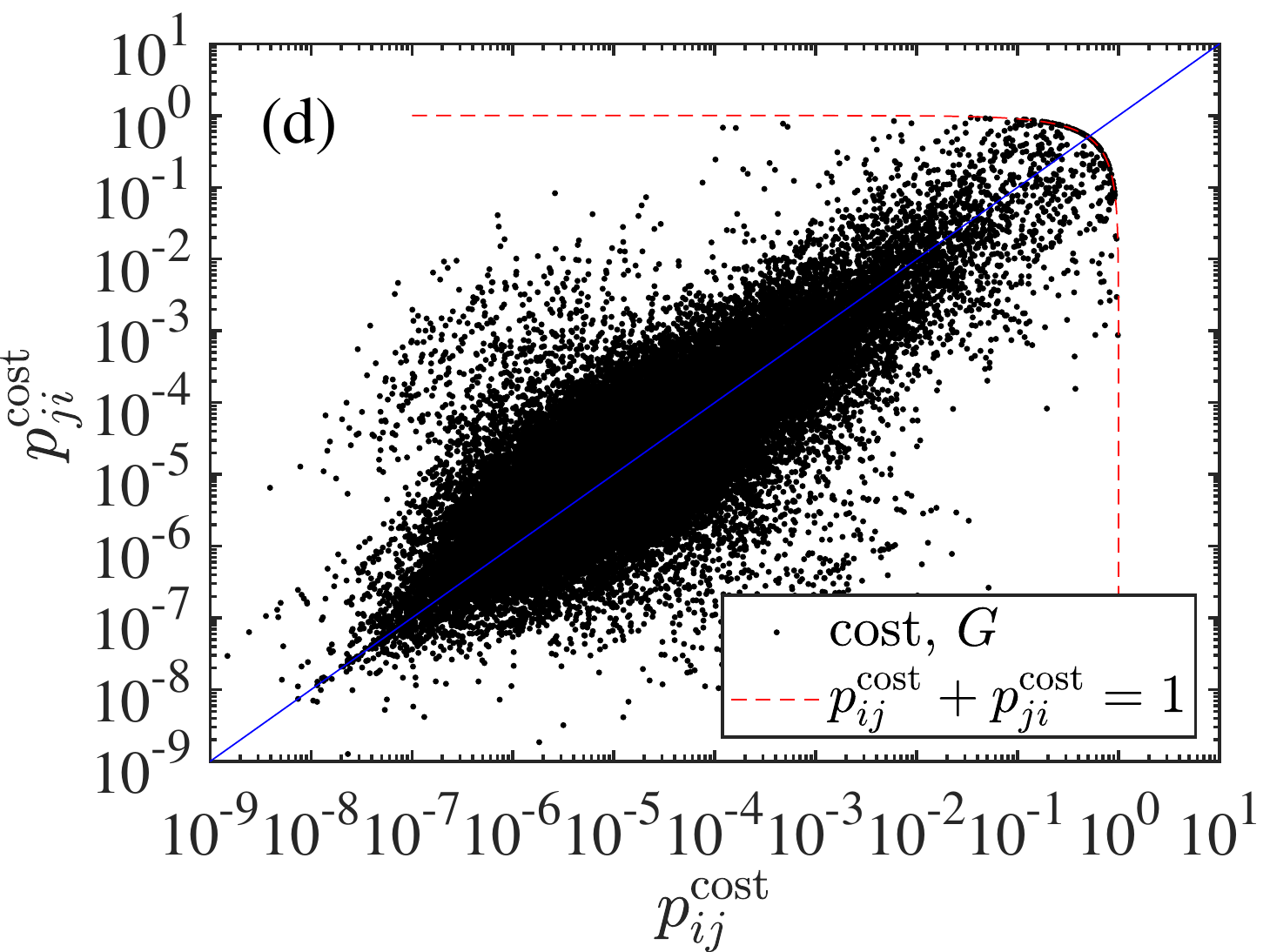}
  \caption{Asymmetric relationship between $p_{ij}$ and $p_{ji}$. (a) Population $P$ is used in the raw radiation model with the geographic distance. (b) Population $P$ is used in the cost-based radiation model with the driving distance. (c) Gross domestic product $G$ is used in the raw radiation model with the geographic distance. (d) Gross domestic product $G$ is used in the cost-based radiation model with the driving distance. }
  \label{Fig:pij:pji}
\end{figure*}

\subsection{Asymmetric relationship between $p_{ij}$ and $p_{ji}$}

We illustrate in Fig.~\ref{Fig:pij:pji} the asymmetric relationship between $p_{ij}$ and $p_{ji}$ for the two radiation models using population. The results for GDP is very similar for each model. It is striking that the predicted values of transportation probability span nine orders of magnitude. We also find that the scatter points lies close to the diagonal $p_{ij}=p_{ji}$. The points from the cost-based model in Fig.~\ref{Fig:pij:pji}(b) concentrate more to the diagonal than the points in Fig.~\ref{Fig:pij:pji}(a) and thus the transportation probability matrix $\{p_{ij}\}$ is less asymmetric. The two dashed lines impose a restriction on the transportation probability values, requiring that
\begin{equation}
  p_{ij}+p_{ji} = 1,
  \label{Eq:Radiation:Pij+pji=1}
\end{equation}
which is more visible if we use linear coordinates. This restriction can be derived as follows.

\begin{figure*}[!t]
  \centering
  \includegraphics[width=0.49\linewidth]{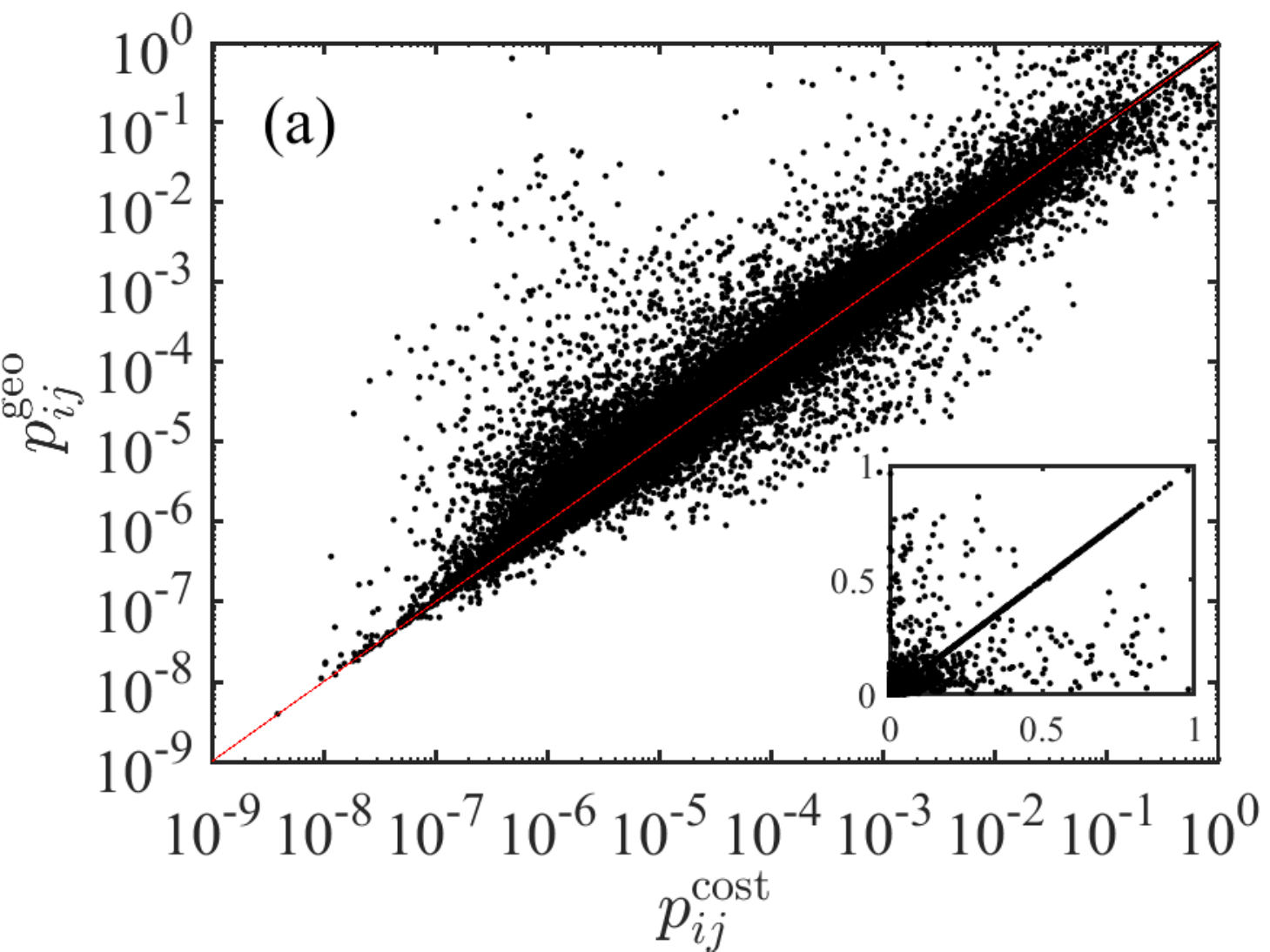}
  \includegraphics[width=0.49\linewidth]{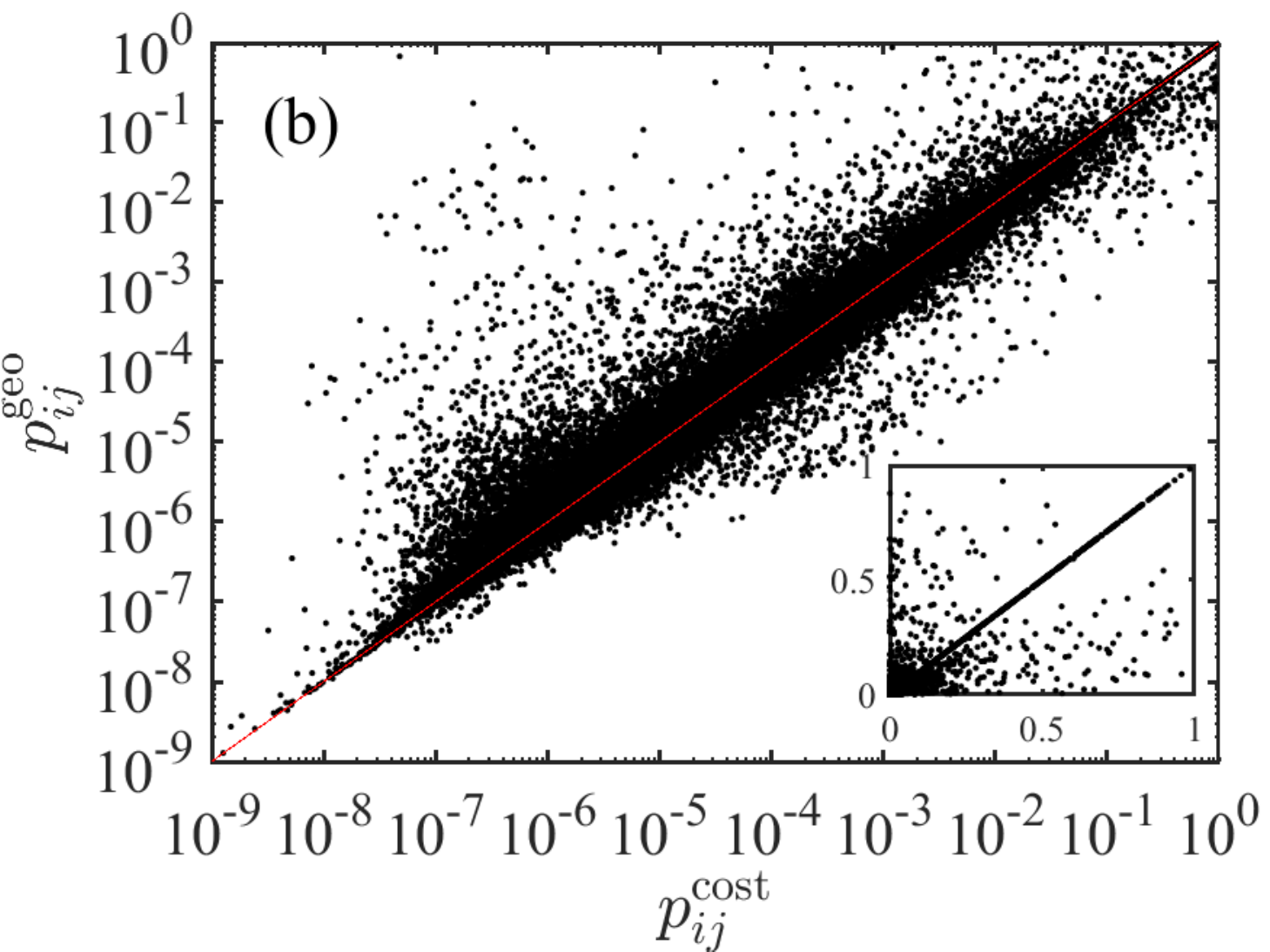}
  \caption{Comparison of the transportation probabilities $p_{ij}$ from the two models based on geographic distance and driving distance. The insets are the same data in linear coordinates. (a) The radiation models are based on population. (b) The radiation models are based on GDP.
  }
  \label{Fig:pij:cost:geo}
\end{figure*}

According to Eq.~(\ref{Eq:Radiation:pij:P}), the probability of transportation from city $j$ to city $i$ is
\begin{equation}
  p_{ji} = \frac{P_iP_j}{(P_j+S_{ji})(P_i+P_j+S_{ji})}.
  \label{Eq:Radiation:pji:P}
\end{equation}
For two given cities $i$ and $j$, it is easy to notice that $p_{ij}$ and $p_{ji}$ reach their maxima when the two cities are adjacent, that is
\begin{equation}
  S_{ij} = S_{ji} = 0.
\end{equation}
In this case, we have
\begin{equation}
  p_{ij} = \frac{P_j}{P_i+P_j}
  \label{Eq:Radiation:pij:P:Sij=0}
\end{equation}
and
\begin{equation}
  p_{ji} = \frac{P_i}{P_i+P_j}.
  \label{Eq:Radiation:pji:P:Sji=0}
\end{equation}
The restriction shown in Eq.~(\ref{Eq:Radiation:Pij+pji=1}) is thus obtained.
This argument holds for both of the radiation models, because the derivation is independent of the definition of the distance between two cities. It also applies to the two models based on GDP, as expressed in Eq.~(\ref{Eq:Radiation:pij:G}).

\subsection{Comparison between $p_{ij}^{\rm{geo}}$ and $p_{ij}^{\rm{cost}}$} 

We compare the predicted transportation probabilities from the two models. The results are shown in Fig.~\ref{Fig:pij:cost:geo}. We find that the points fluctuate around the diagonal line
\begin{equation}
  p_{ij}^{\rm{cost}} = p_{ij}^{\rm{geo}}.
  \label{Eq:pgeo:gt:pcost}
\end{equation}
The insets show that there are many points that fall exactly on the diagonal. These points correspond to the situations when 
\begin{equation}
  S_{ij}^{\rm{geo}} = S_{ij}^{\rm{cost}}.
  \label{Eq:Sij:geo:cost}
\end{equation}
Usually, this condition (\ref{Eq:Sij:geo:cost}) is more likely to be fulfilled when the two cities $i$ and $j$ are close. As a special case, when city $j$ is the closest city of city $i$, we have $S_{ij}^{\rm{geo}}=S_{ij}^{\rm{cost}}=0$. In this case, the two transportation probabilities $p_{ij}^{\rm{geo}}$ and $p_{ij}^{\rm{cost}}$ are identical.

\section{Transportation diversity}
\label{S1:Diversity}

We now define the transportation diversity of a city $i$ based on its transportation probability $p_{ij}$ as follows
\begin{equation}
  D_{i} = -\sum_{i\neq{j}} p_{ij}\ln p_{ij},
  \label{Eq:Radiation:Diversity}
\end{equation}
where $p_{ij}$ can be calculated from the two radiation models using either population $P$ or gross domestic product $G$. We calculate four sets of diversity $D_{i}^{M, d}$, where $M=P$ or $M=G$ and $d=d^{\rm{geo}}$ or $d=d^{\rm{cost}}$. Indeed, human mobility or communication diversity has been proposed and studied \cite{Eagle-Macy-Claxton-2010-Science,Yan-Han-Wang-Zhou-2013-SR,Liu-Yan-2020-SciRep}.

\subsection{Comparison of diversity based on population and gross domestic product}

\begin{figure*}[!t]
  \centering
  \includegraphics[width=0.47\linewidth]{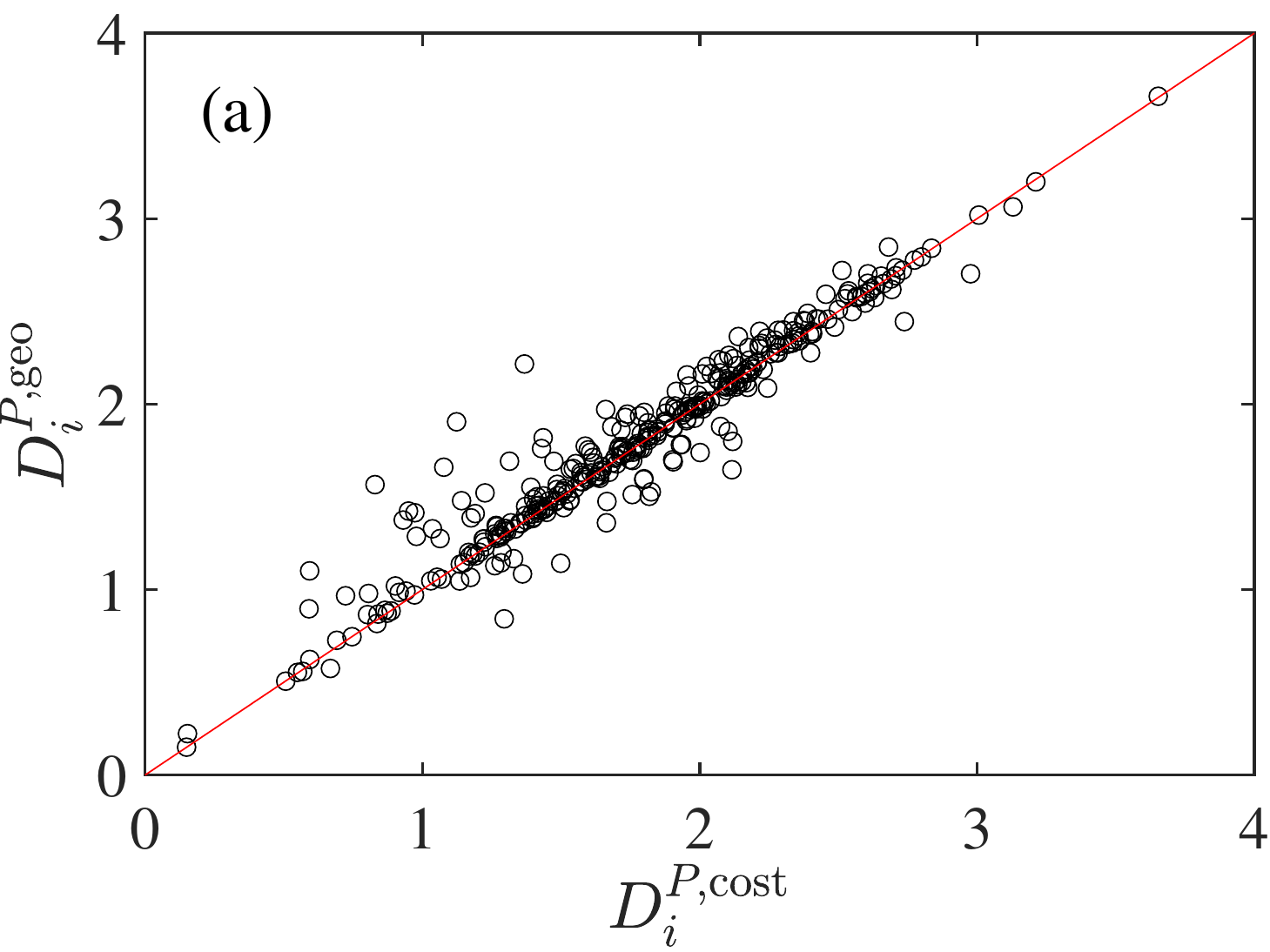}
  \includegraphics[width=0.47\linewidth]{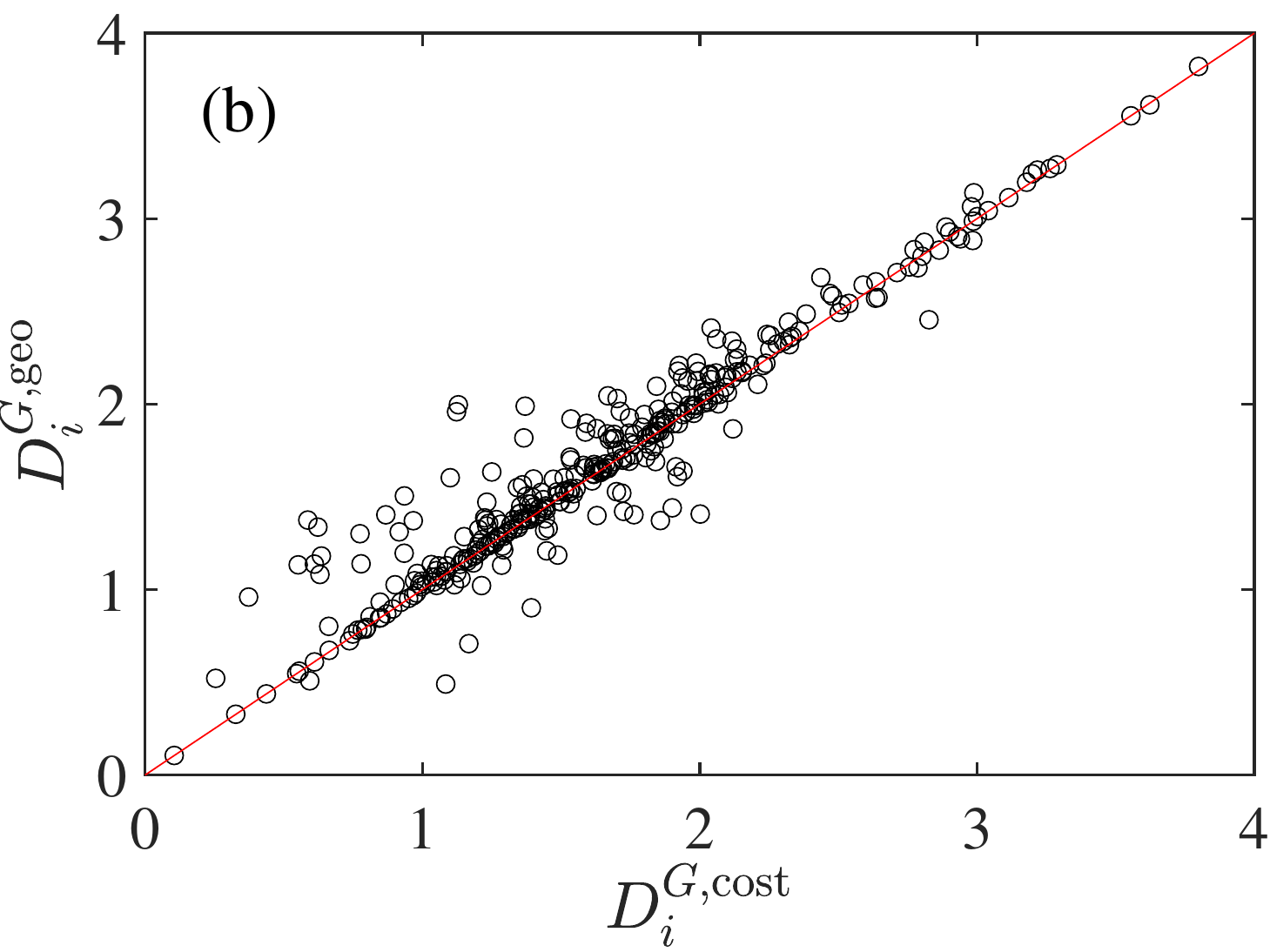}\\
  \includegraphics[width=0.47\linewidth]{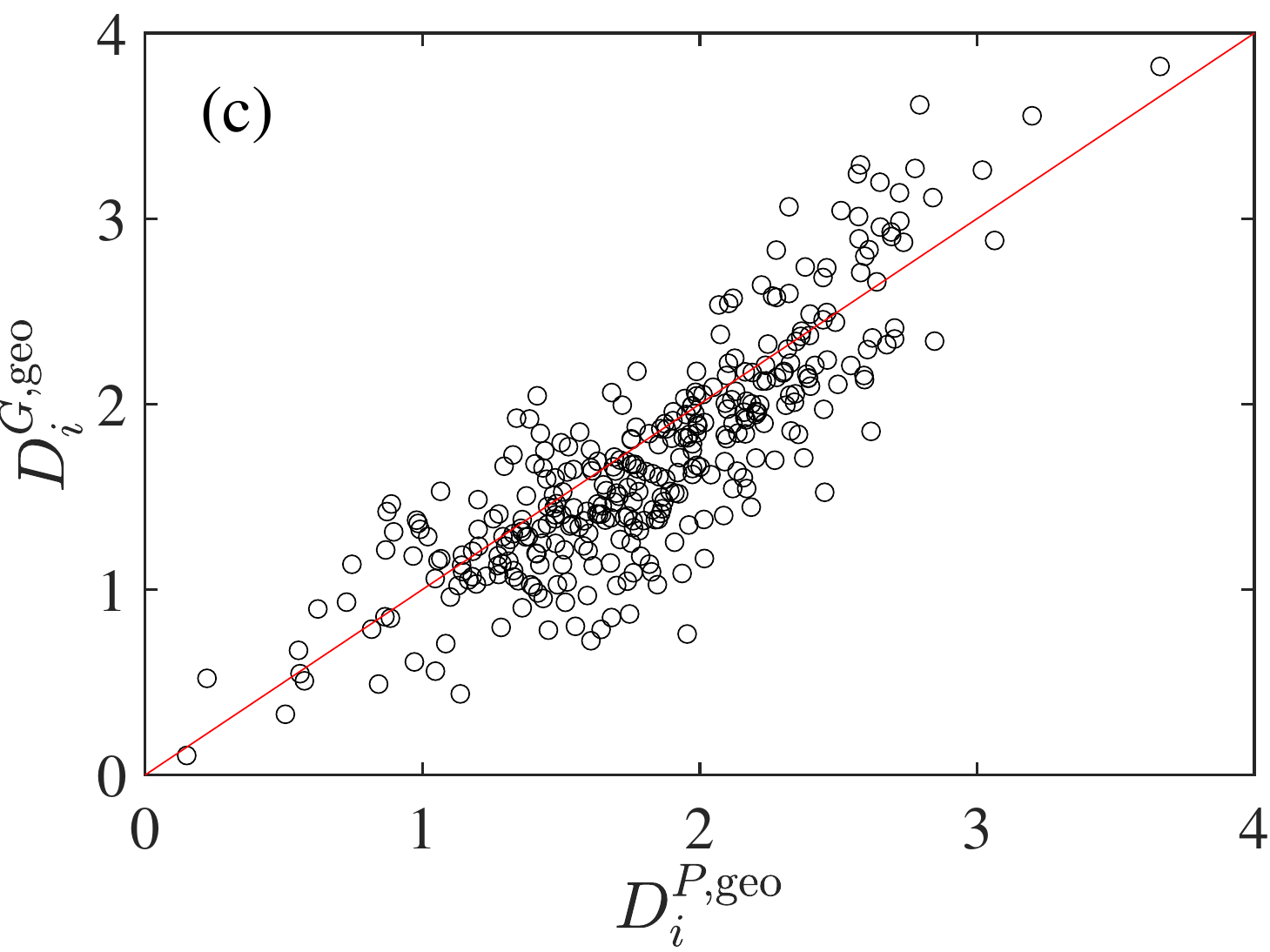}
  \includegraphics[width=0.47\linewidth]{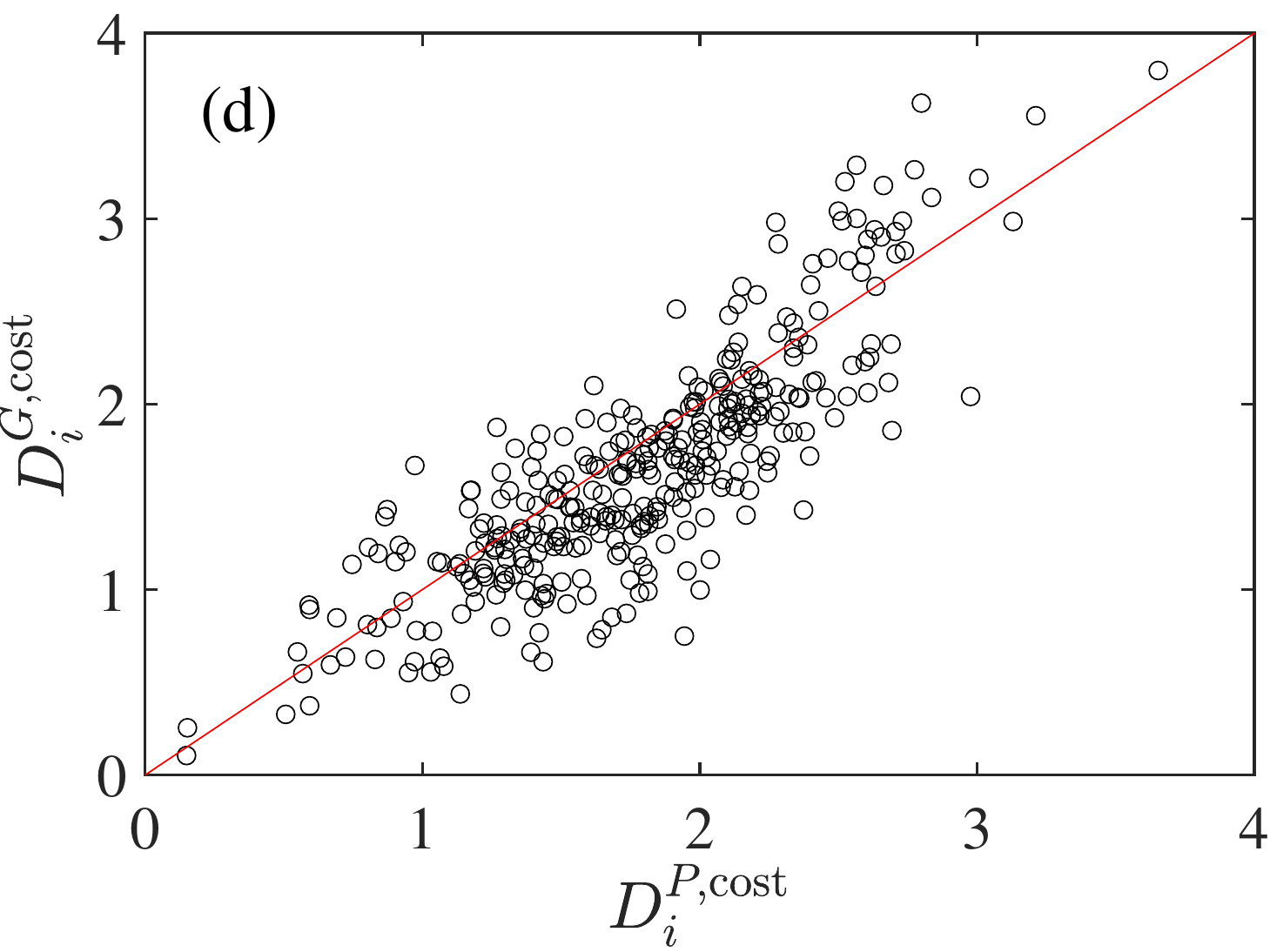}  
  \caption{Comparison of the two transportation diversity measures $D_i^{M^{(1)},d^{(1)}}$ and $D_i^{M^{(2)},d^{(2)}}$ calculated using population $P$ and gross domestic product $G$ for the raw radiation model and the cost-based radiation model. 
  (a) $M^{(1)}=M^{(2)}=P$, $d^{(1)}=d^{\rm{geo}}$ and $d^{(2)}=d^{\rm{cost}}$. 
  (b) $M^{(1)}=M^{(2)}=G$, $d^{(1)}=d^{\rm{geo}}$ and $d^{(2)}=d^{\rm{cost}}$. 
  (c) $d^{(1)}=d^{(2)}=d^{\rm{geo}}$, $M^{(1)}=G$, and $M^{(2)}=P$. 
  (d) $d^{(1)}=d^{(2)}=d^{\rm{cost}}$, $M^{(1)}=G$, and $M^{(2)}=P$. 
  The solid lines are the diagonal lines.}
  \label{Fig:Di:Di}
\end{figure*}

In Fig.~\ref{Fig:Di:Di}, we compare six pairs of any two diversity sets obtained. The two plots in the top row show the influence of distance on diversity for fixed choice of $M$, while the two plots in the bottom row illustrate the influence of the choice of $M$ on diversity in a given model. We find that, in each plot, there is a nice linear relationship:
\begin{equation}
  D_i^{M^{(1)},d^{(1)}} = D_i^{M^{(2)},d^{(2)}}.
\end{equation}
It is found that the influence is weaker for the choice of model than for the choice of $M$.

\subsection{Dependence of city traits on diversity}

\begin{figure*}[!ht]
  \centering
  \includegraphics[width=0.24\linewidth]{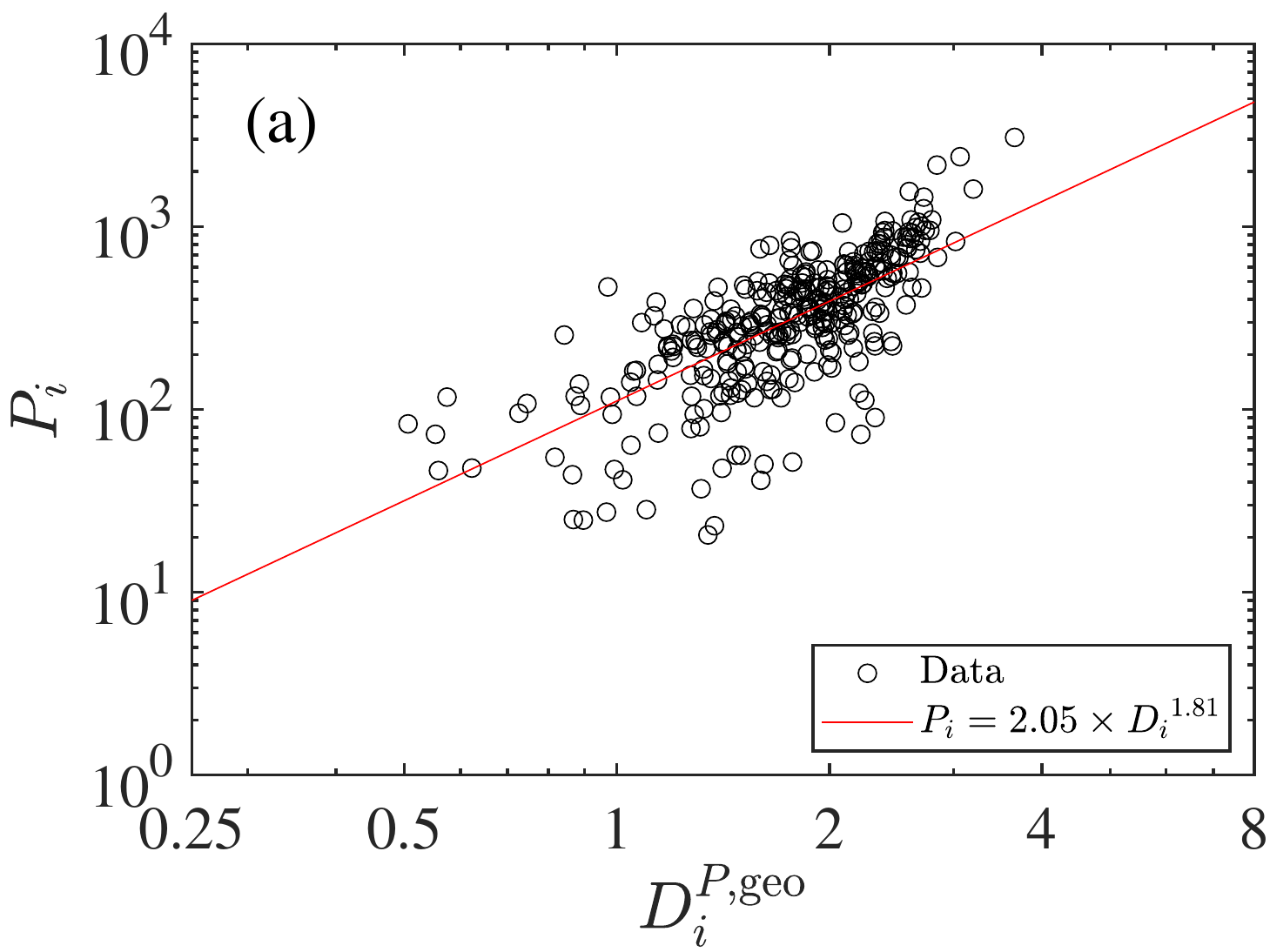}
  \includegraphics[width=0.24\linewidth]{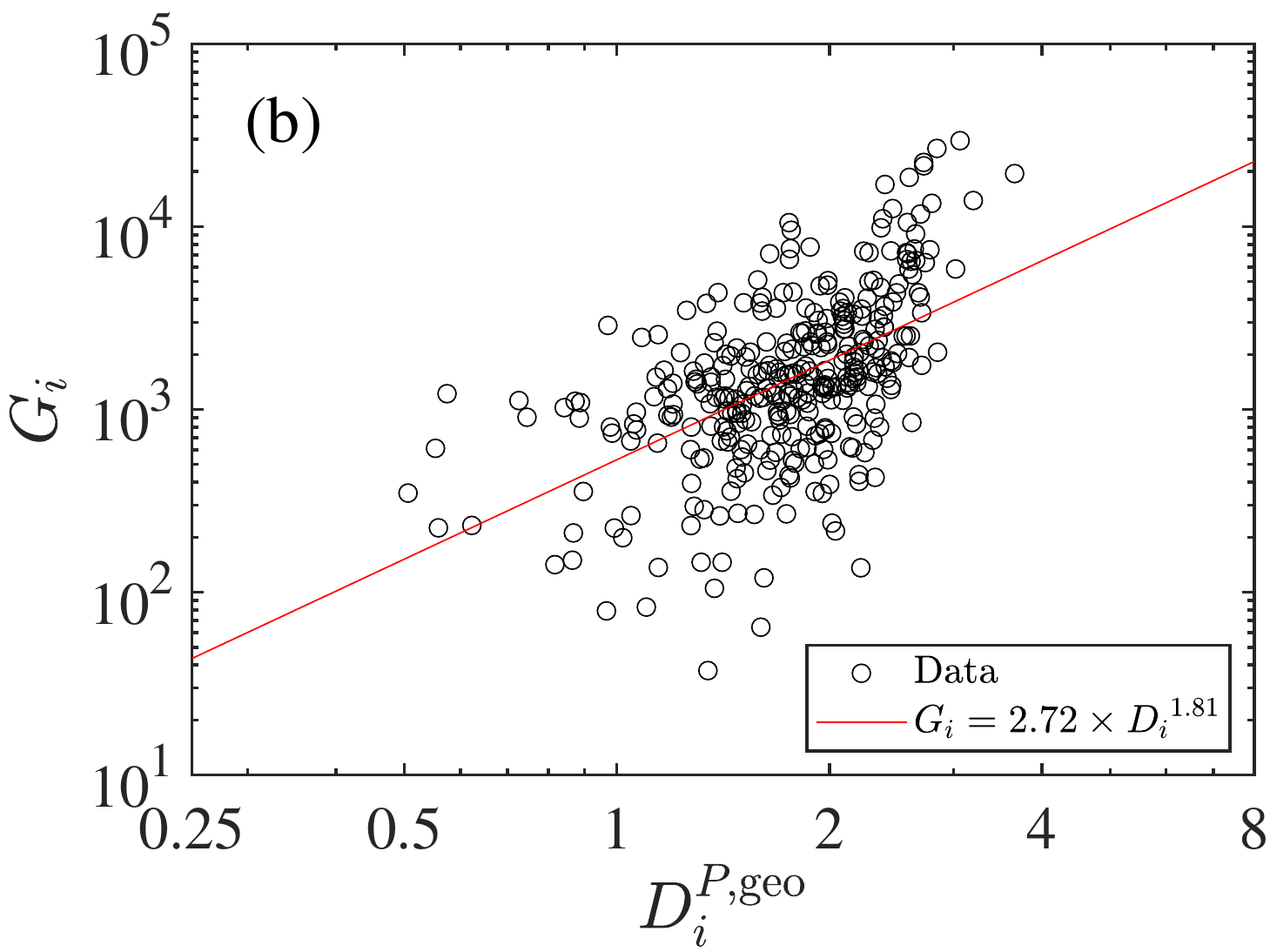}
  \includegraphics[width=0.24\linewidth]{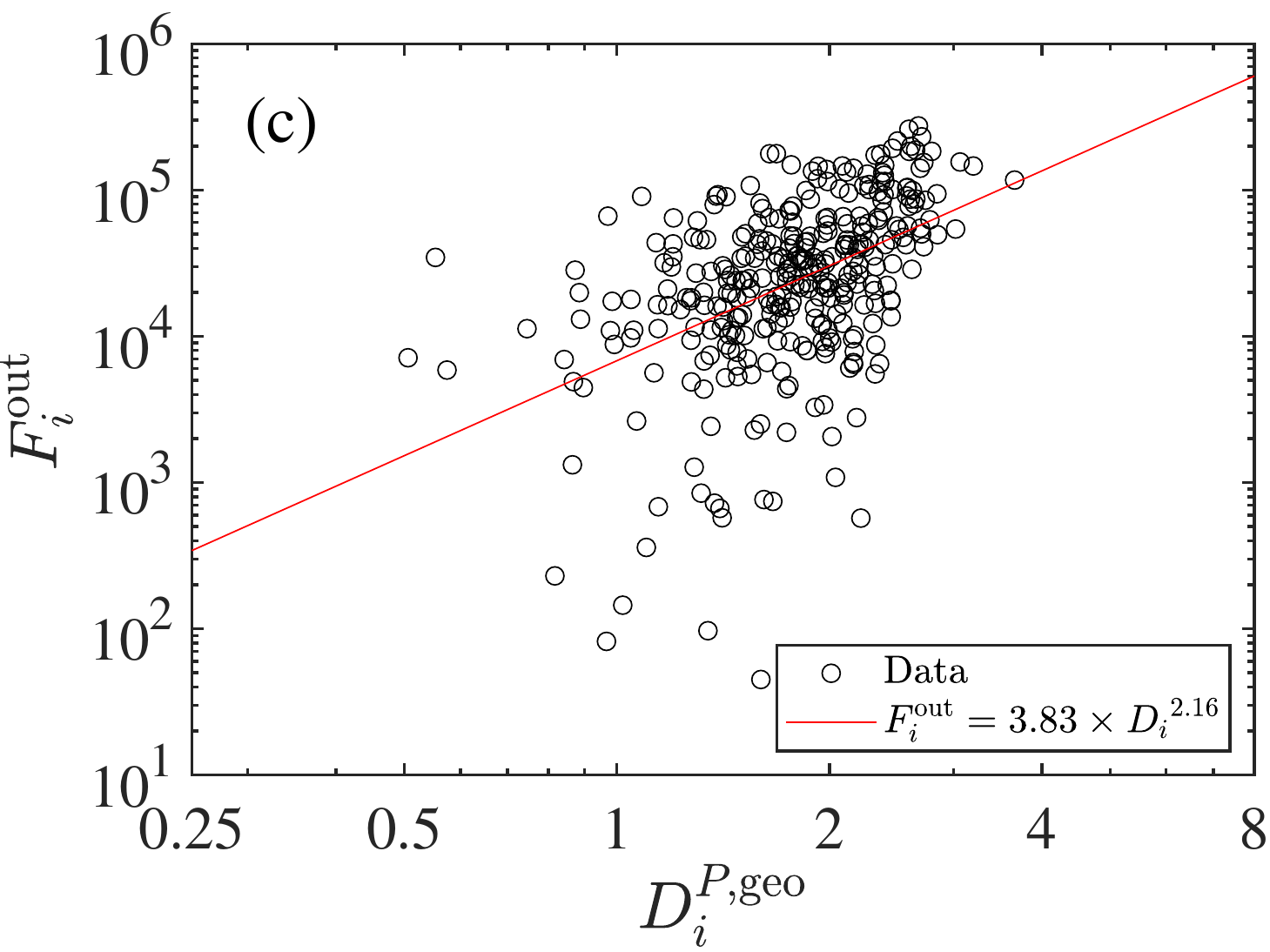}
  \includegraphics[width=0.24\linewidth]{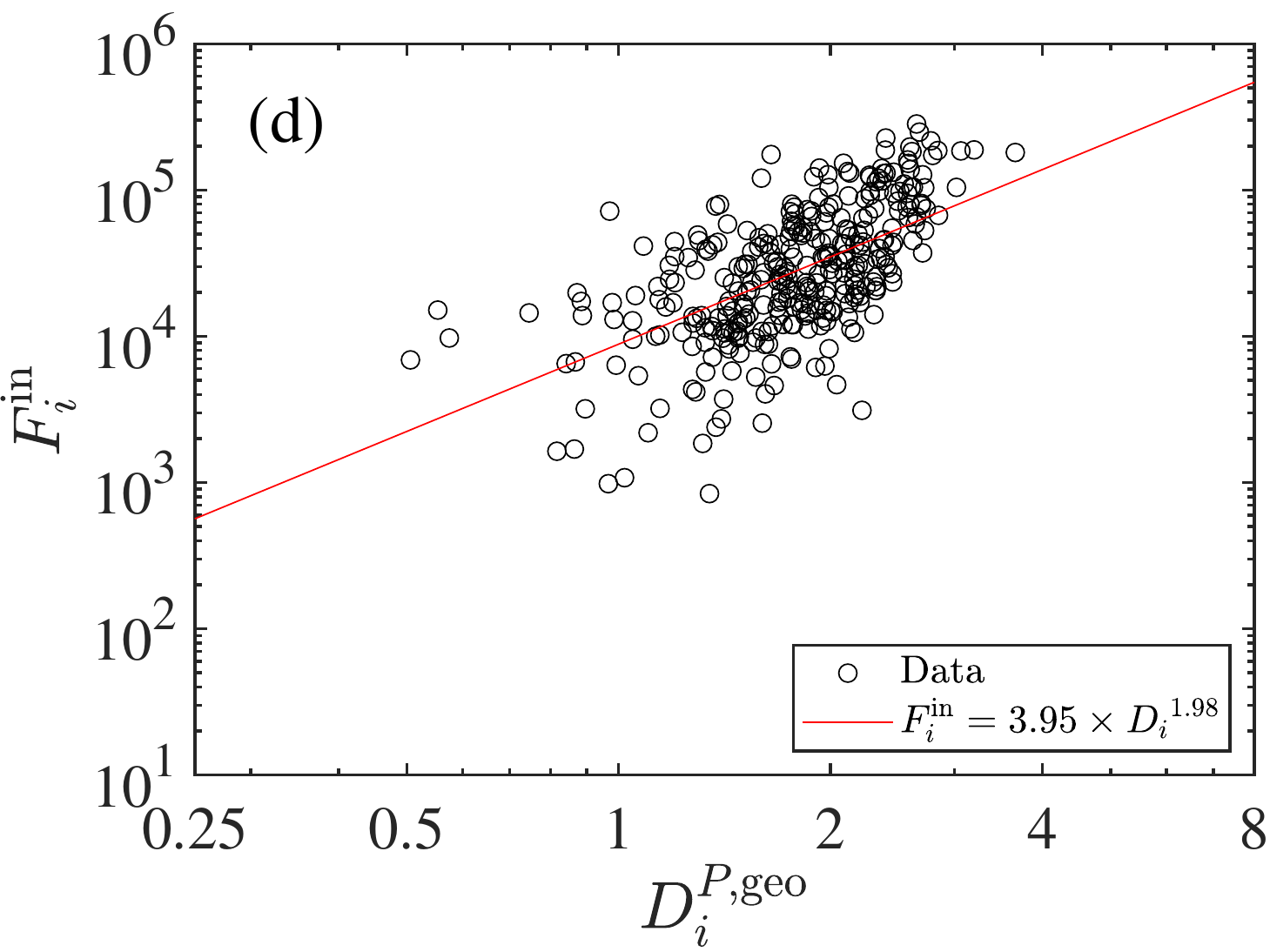}\\
  \includegraphics[width=0.24\linewidth]{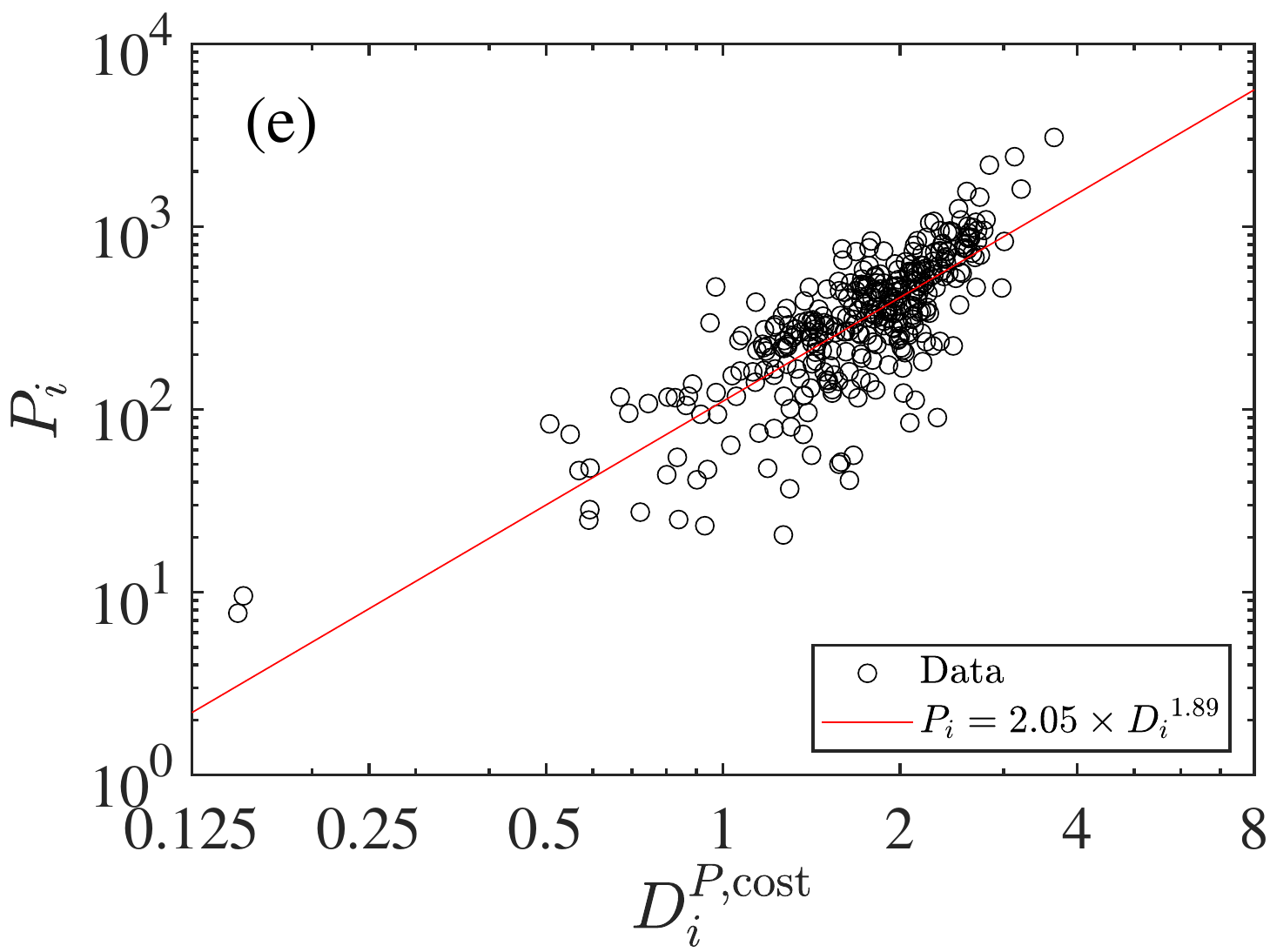}
  \includegraphics[width=0.24\linewidth]{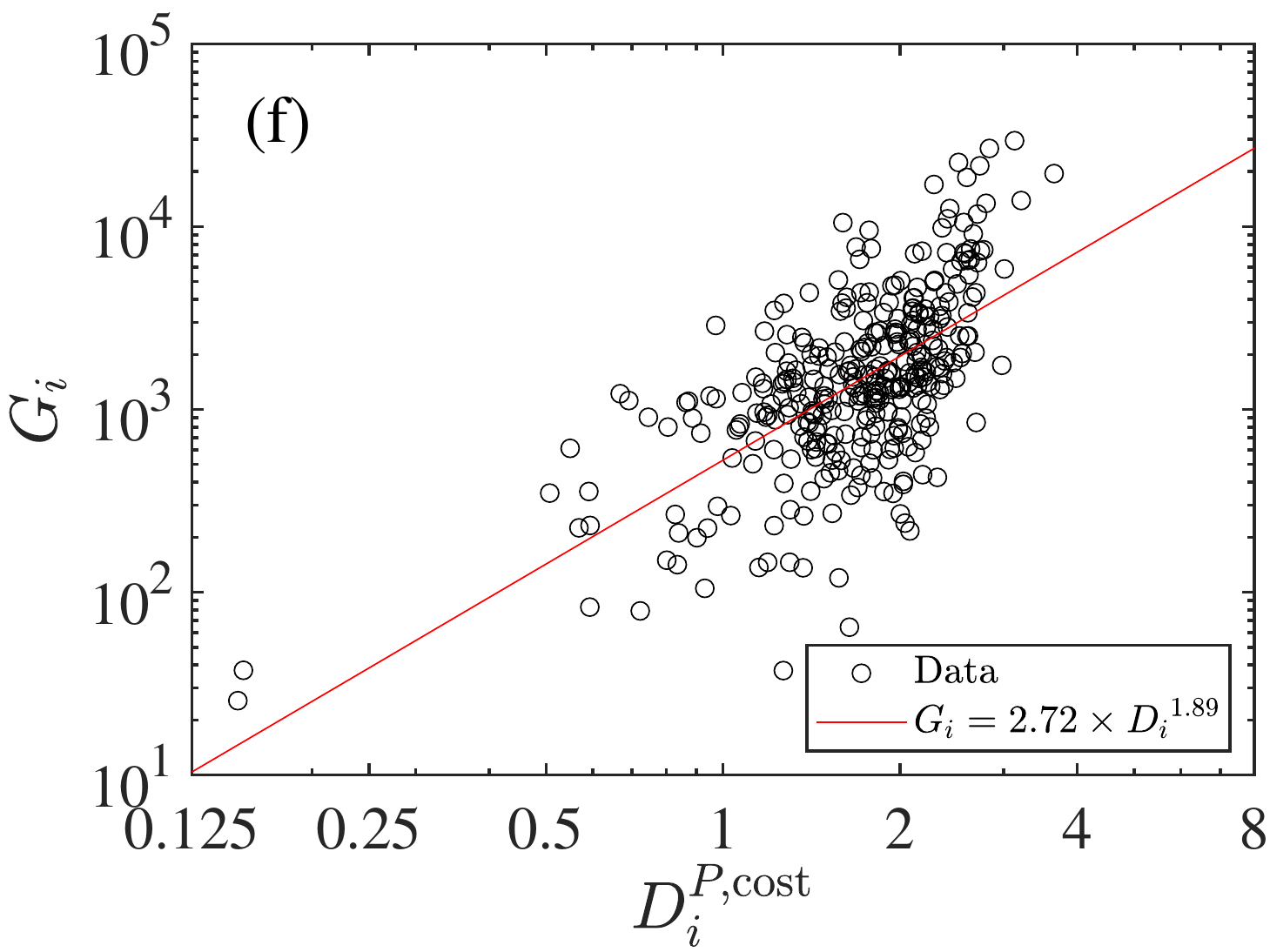}
  \includegraphics[width=0.24\linewidth]{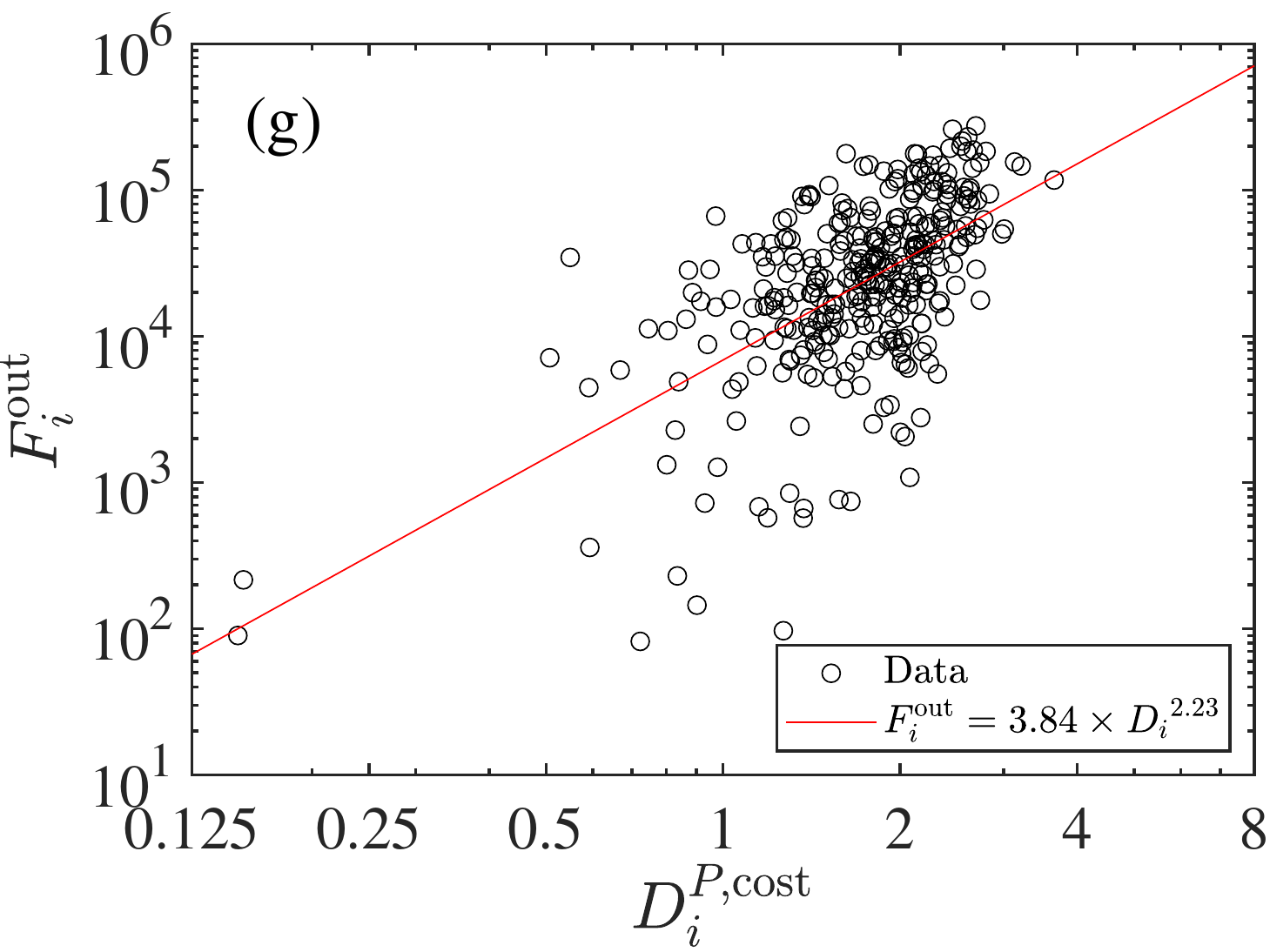}
  \includegraphics[width=0.24\linewidth]{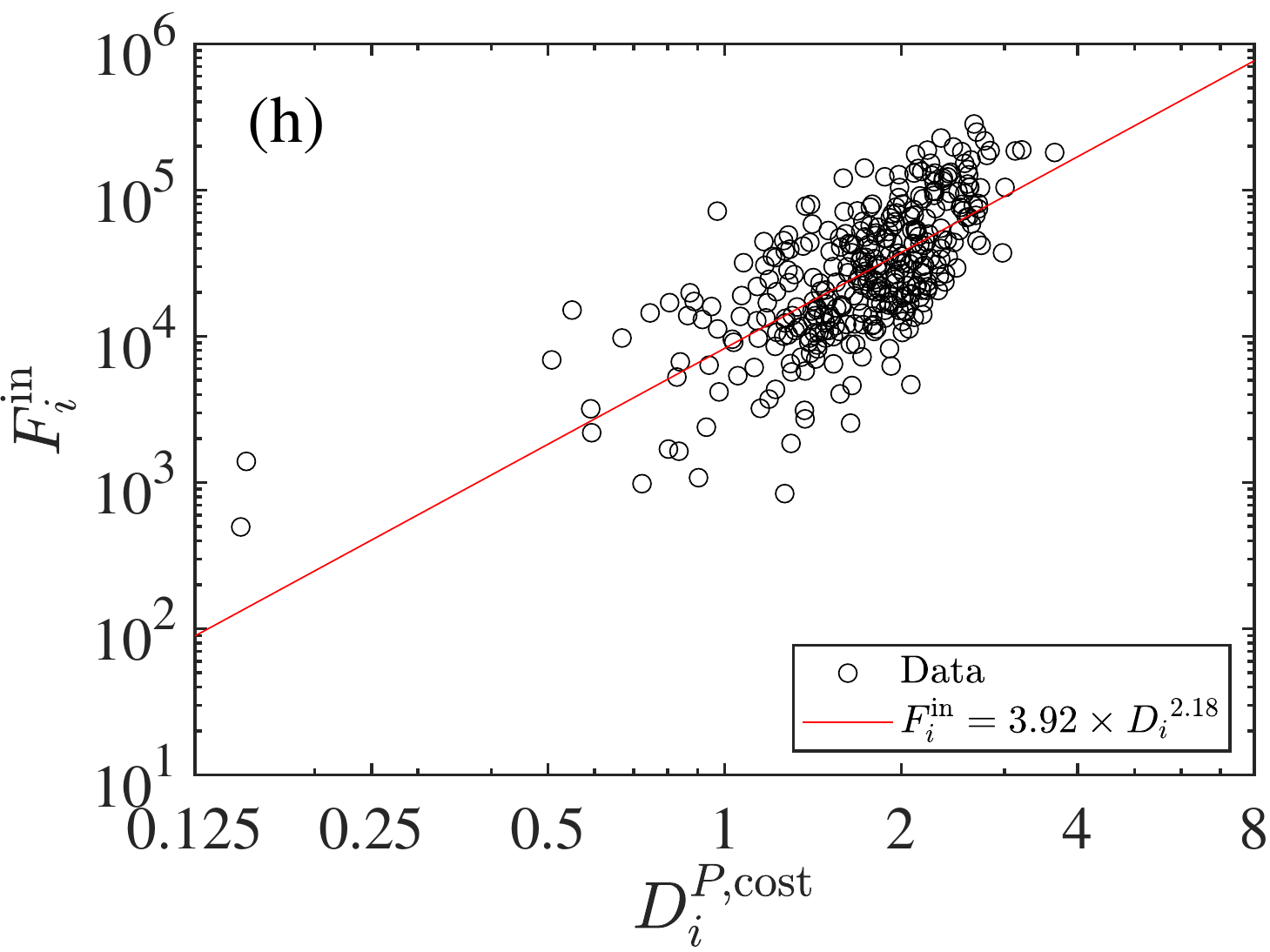}\\
  \includegraphics[width=0.24\linewidth]{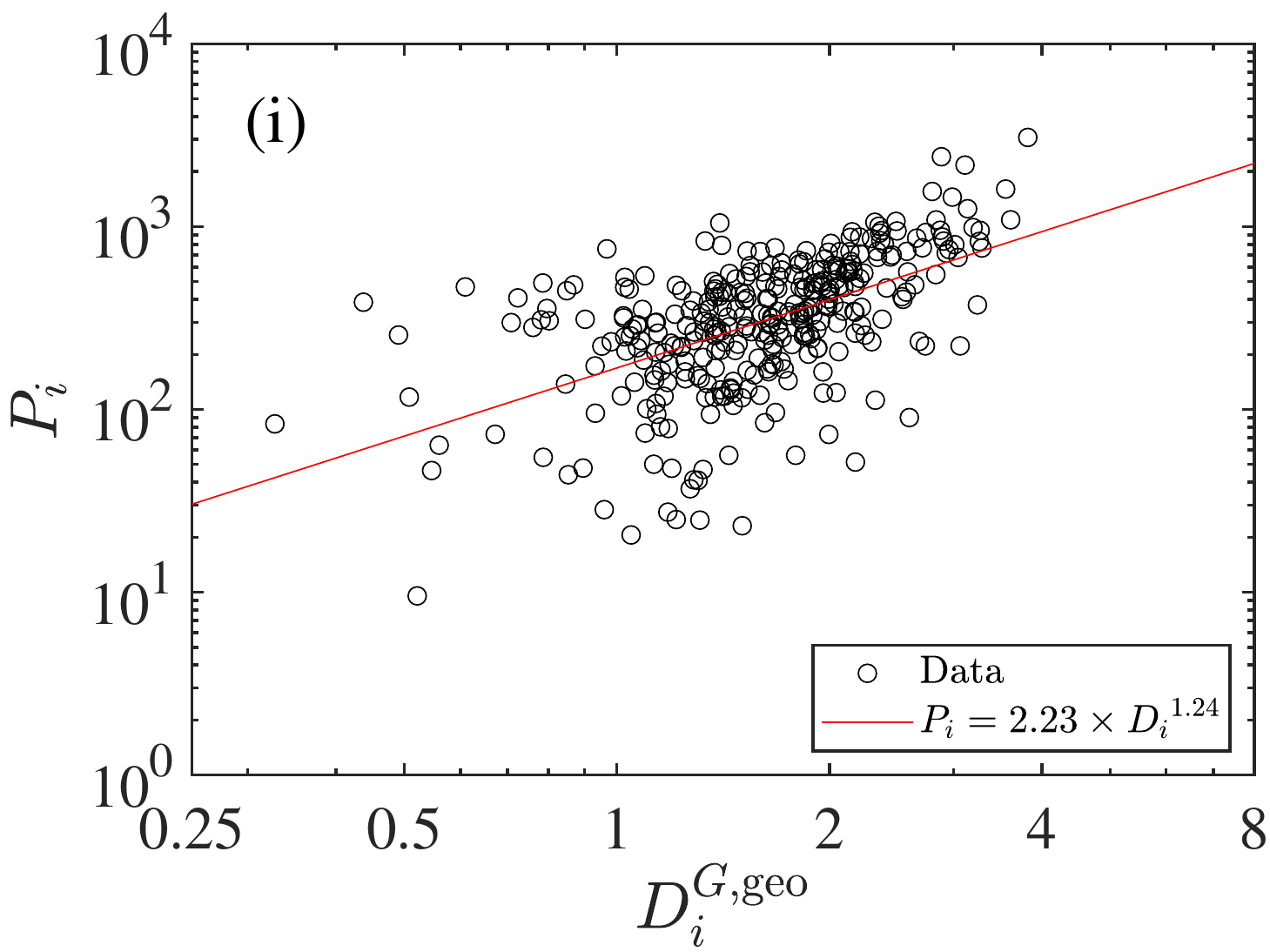}
  \includegraphics[width=0.24\linewidth]{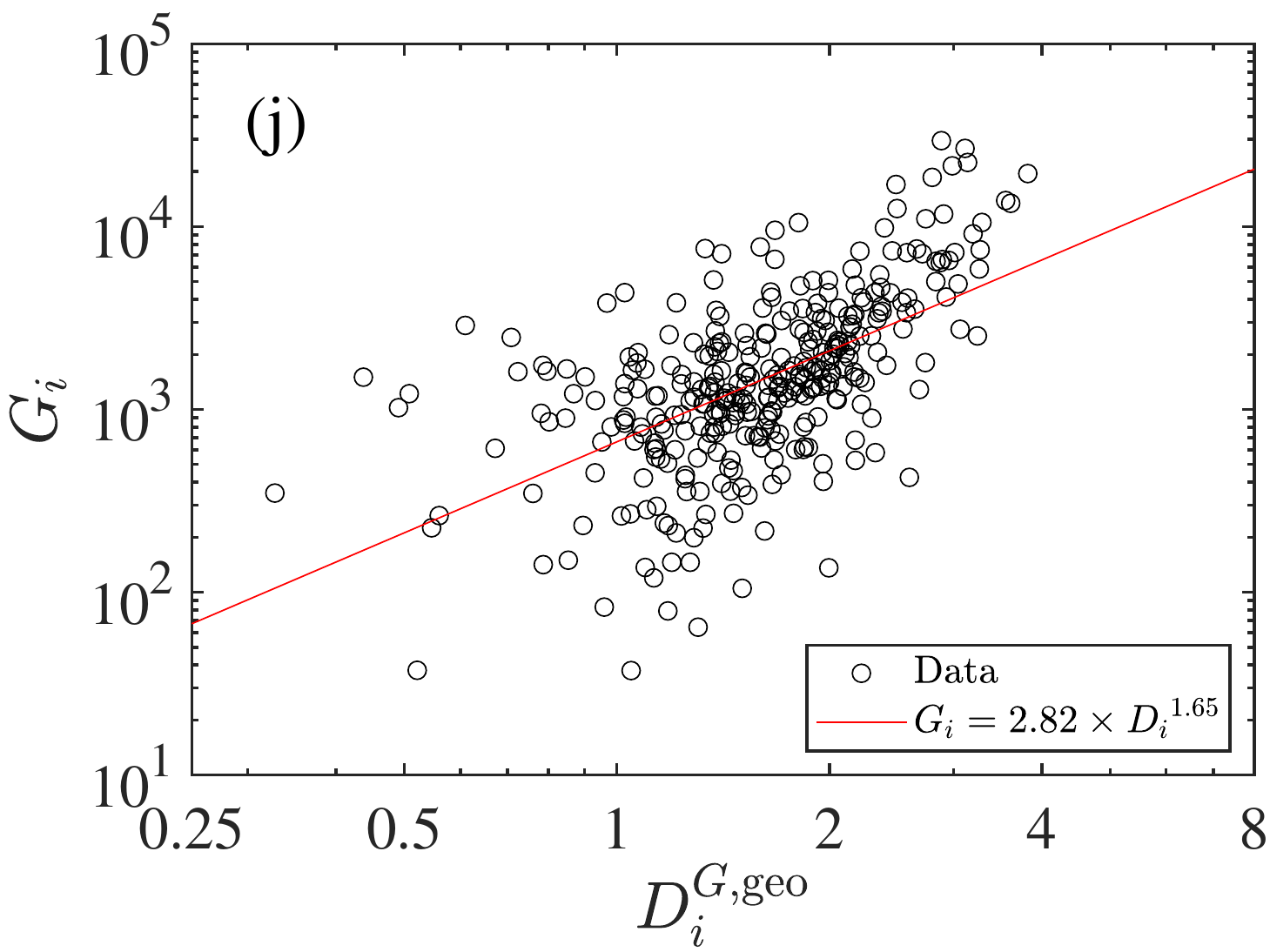}
  \includegraphics[width=0.24\linewidth]{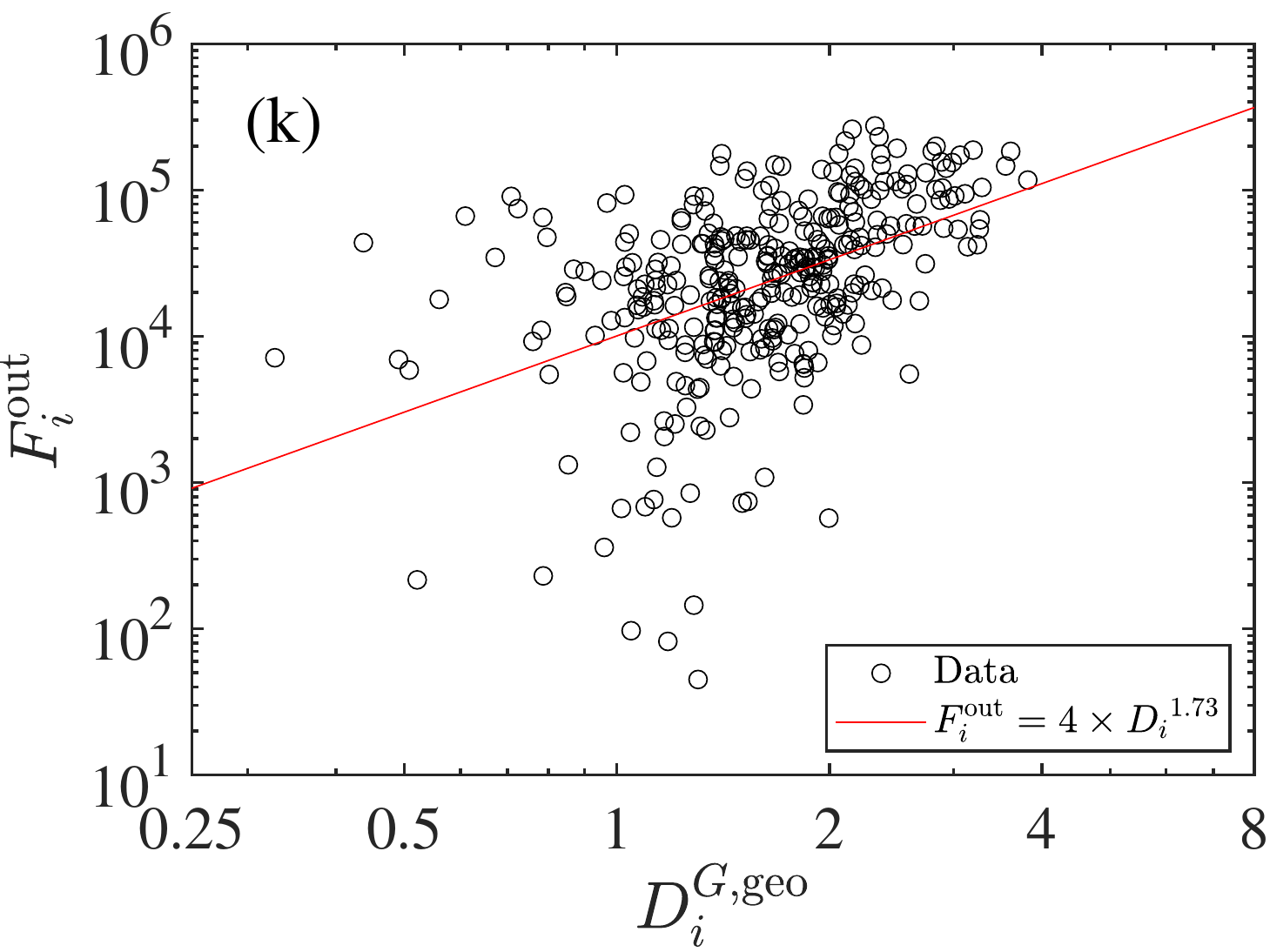}
  \includegraphics[width=0.24\linewidth]{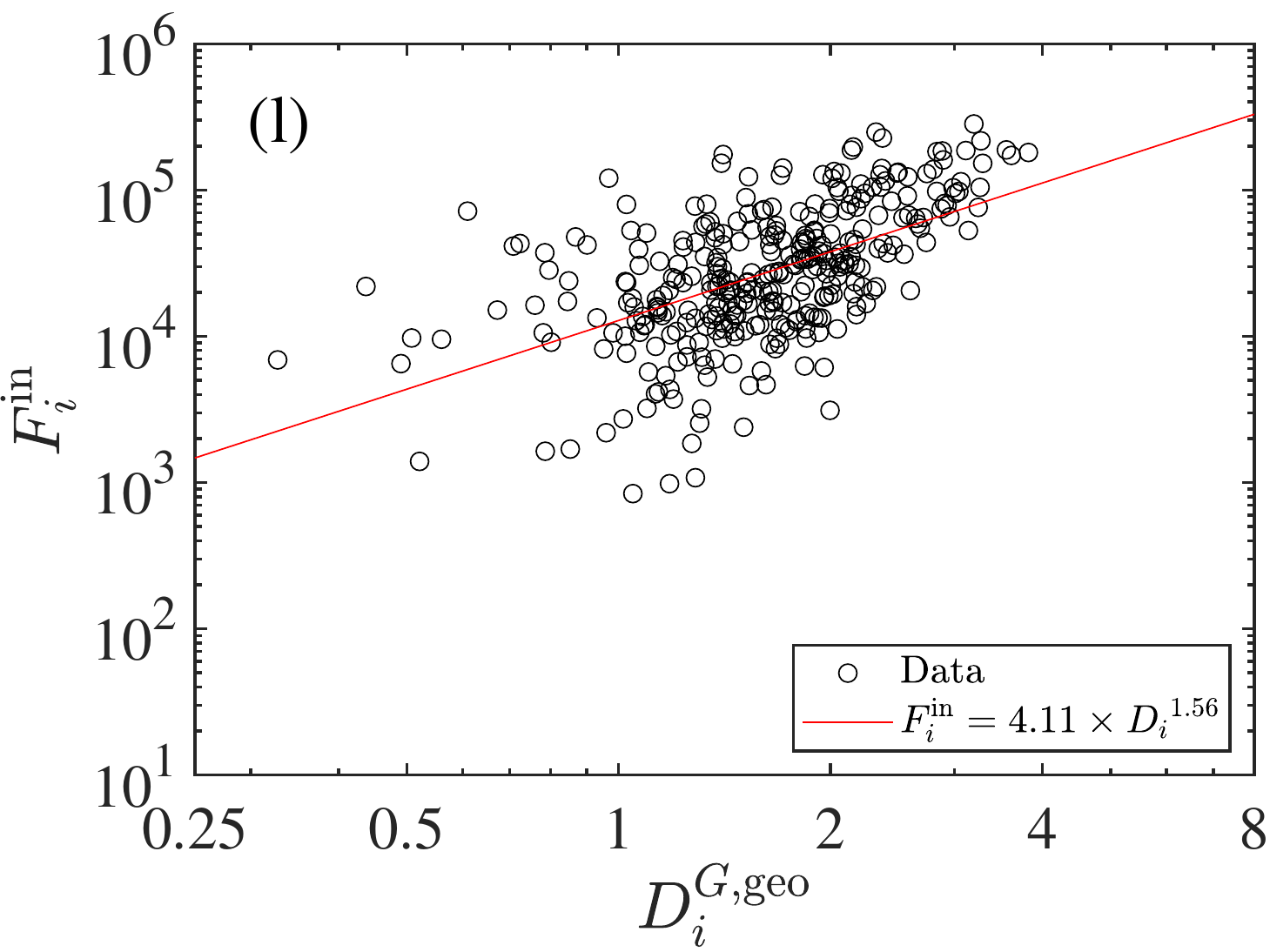}\\
  \includegraphics[width=0.24\linewidth]{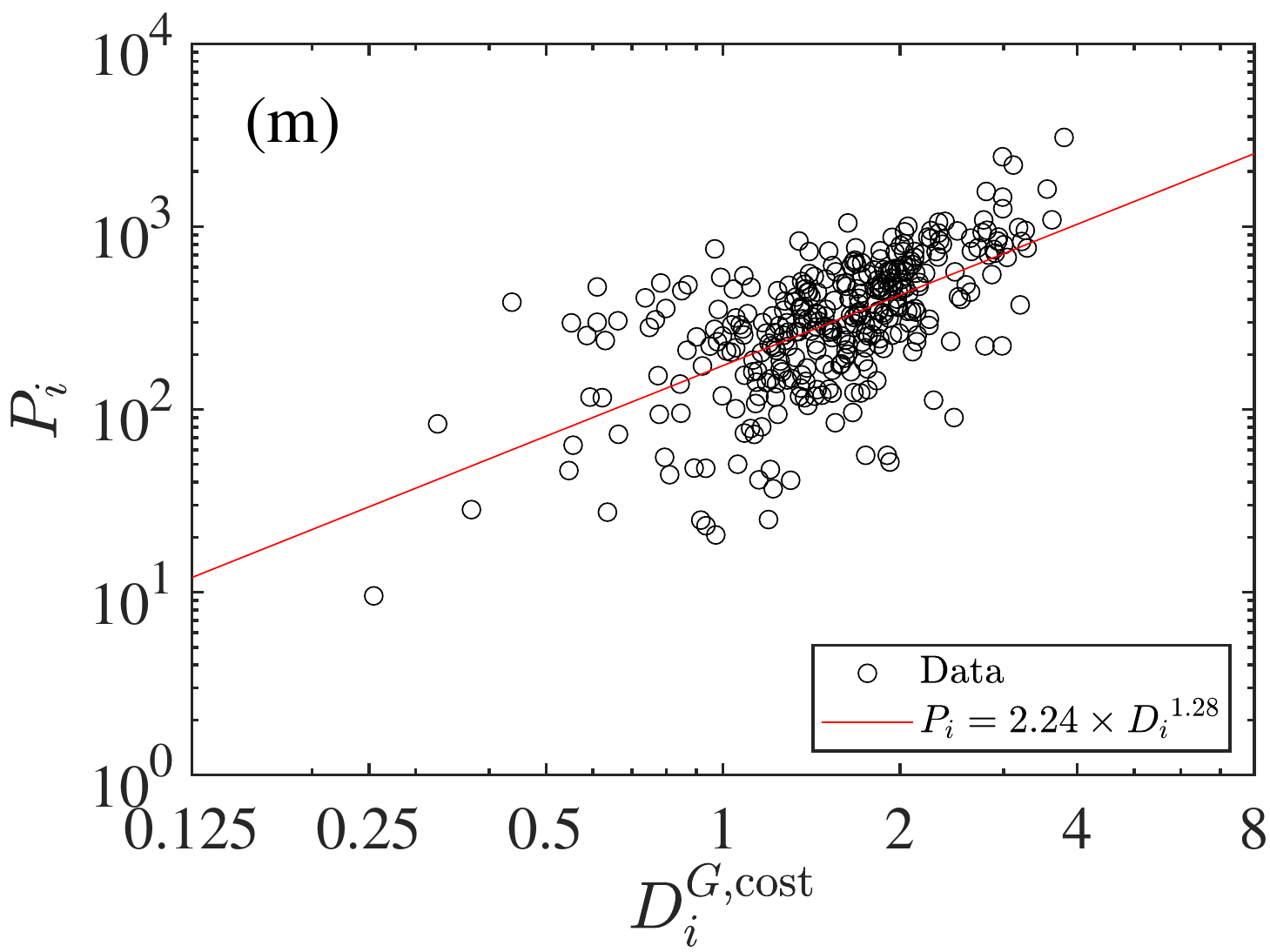}
  \includegraphics[width=0.24\linewidth]{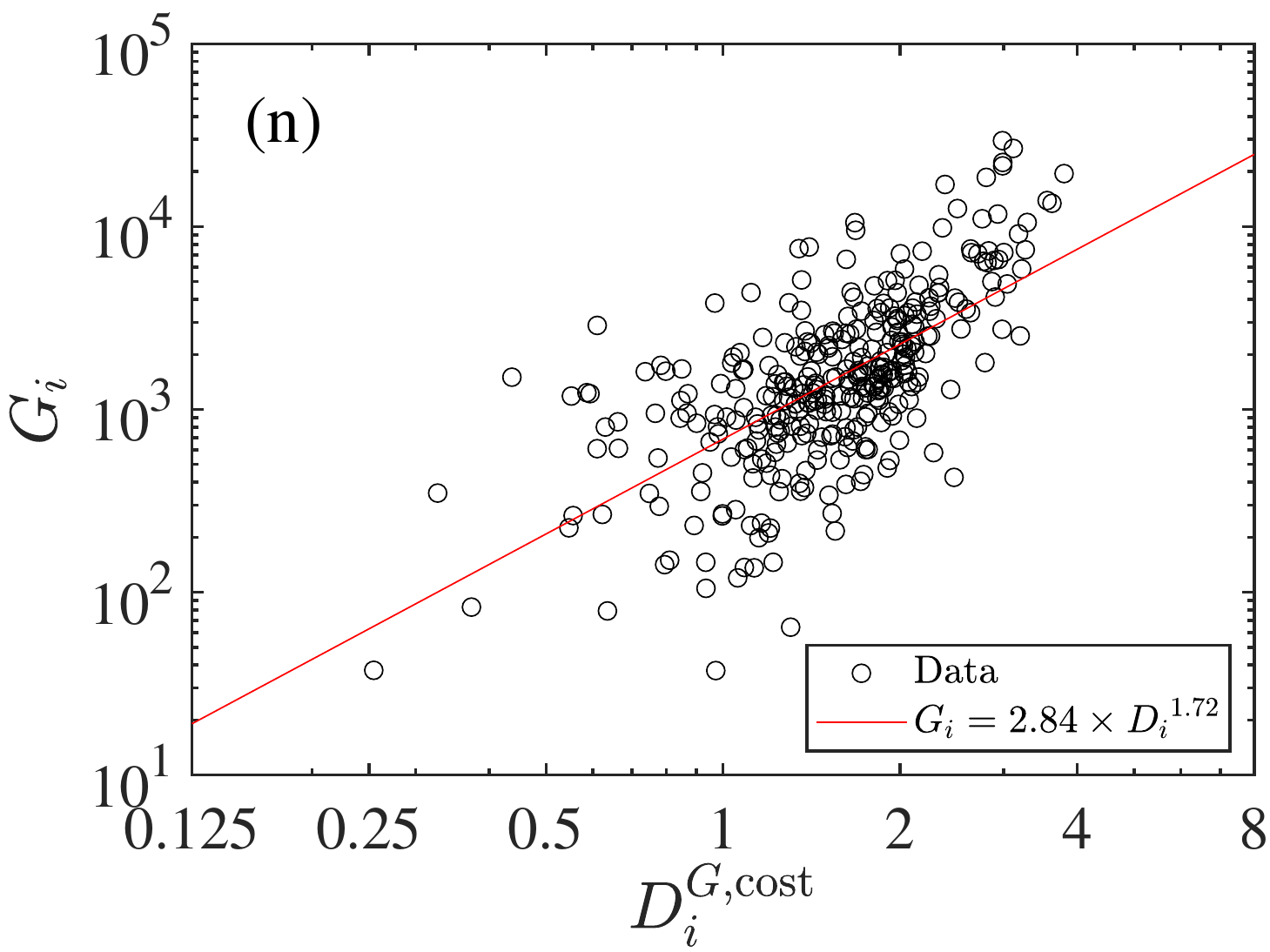}
  \includegraphics[width=0.24\linewidth]{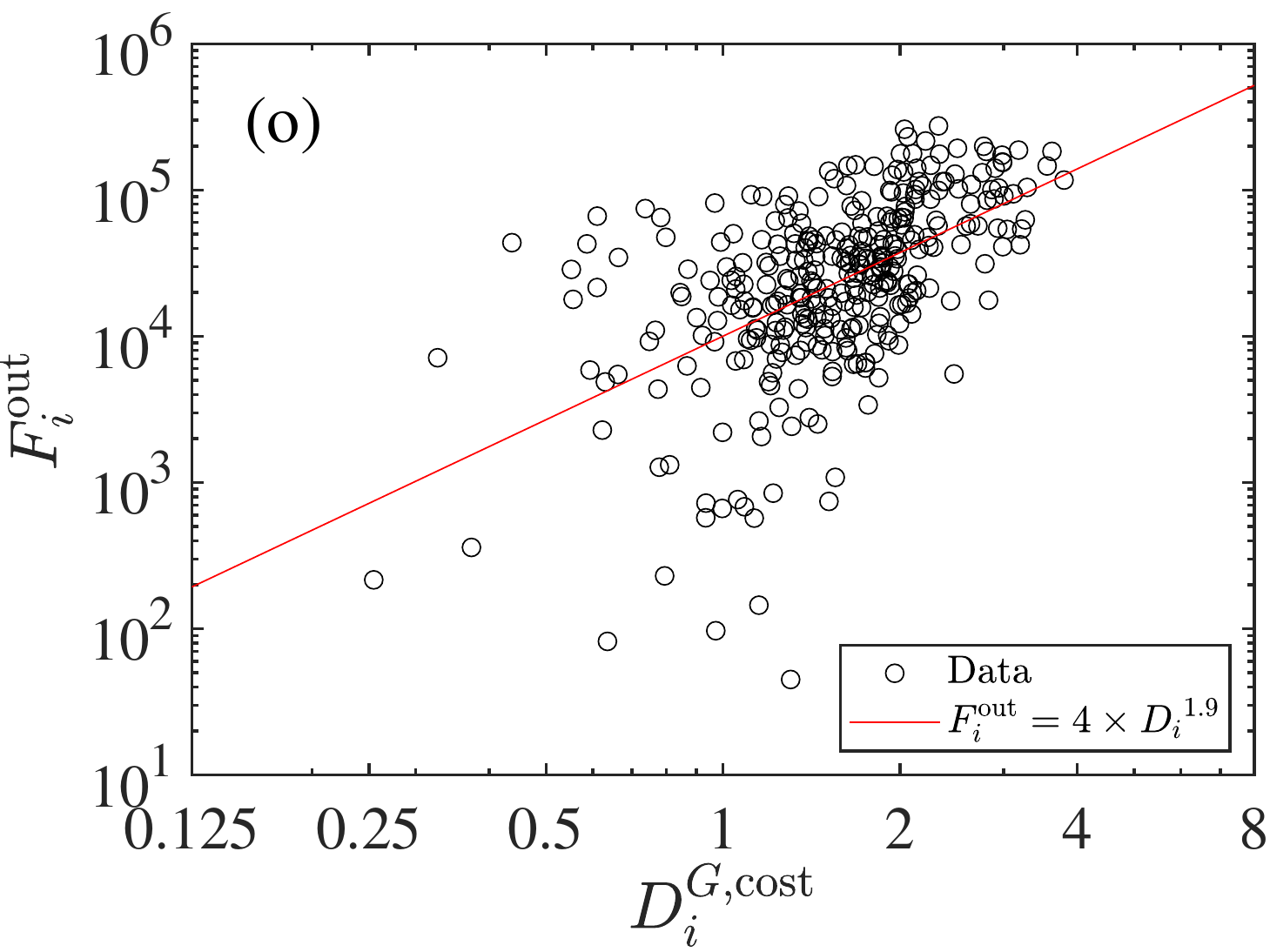}
  \includegraphics[width=0.24\linewidth]{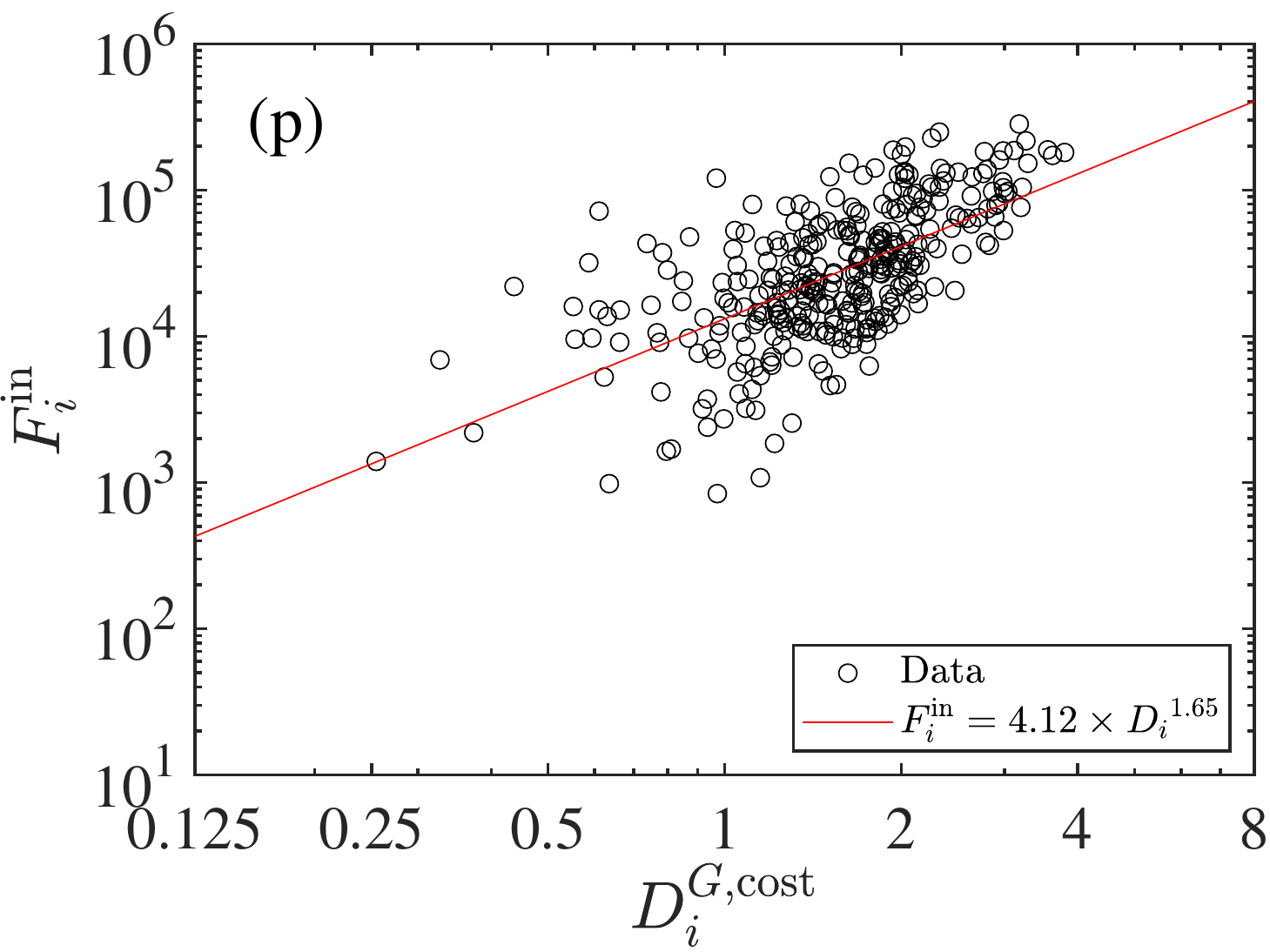}
  \caption{Dependence of city traits ($P$, $G$, $F^{\rm{out}}$, and $F^{\rm{in}}$) on truck transportation diversity ($D^{P,{\rm{geo}}}$). The diversity is calculated from the raw radiation model based on population and geographic distance. The solid lines are power-law fits.}
  \label{Fig:Di:CityTraits}
\end{figure*}

We further check the dependence of city traits ($P$, $G$, $F^{\rm{out}}$, or $F^{\rm{in}}$) on the truck transportation diversity $D_i$, where $F_i^{\rm{in}}$ is total in-flux arriving at city $i$
\begin{equation}
  F_i^{\rm{in}} = \sum_{j\neq{i}} F_{ji}.
\end{equation}
The results are depicted in Fig.~\ref{Fig:Di:CityTraits}. In the four plots of Fig.~\ref{Fig:Di:CityTraits}(e-h) for $D_i^{P,{\rm{cost}}}$, we observe two outliers that seem isolated from other points. These outliers correspond to two same cities, Shennongjia Forestry District and Ali District. The diversities of these two cities are respectively 0.1496 and 0.1529.

We observe power-law dependence in each plot. We can write that
\begin{equation}
  Y_i \sim \left(D_i^{M,d}\right)^{\beta(Y,M,d)},
\end{equation}
where $Y$ represents $P$, $G$, $F^{\rm{out}}$ or $F^{\rm{in}}$, $M$ stands for population $P$ or gross domestic product $G$ in the radiation model, and $d$ determines the geographic or driving distance. The power-law exponents $\beta(Y,M,d)$ are estimated with the ordinary least-squares regression, which are presented in Table~\ref{TB:Radiation:Exponents}. For a given city trait and the chosen $M$, the two power-law exponents are similar in the raw radiation model and the cost-based radiation model. In contrast, the power-law exponent is larger when we use population $P$ as $M$ in the radiation models.

\begin{table}[!ht]
  \centering
  \caption{Power-law exponents $\beta(Y,M,d)$ for the cost-based radiation model.}
  \medskip
  \begin{tabular}{ccccc}
    \hline
    Model & $Y=P$ & $Y=G$ & $Y=F^{\rm{out}}$ & $Y=F^{\rm{in}}$\\
    \hline
    $d^{\rm{geo}}, P$ & 1.8111  &  1.8063  &  2.1558  &  1.9829\\
    $d^{\rm{cost}},P$ & 1.8863  &  1.8890  &  2.2277  &  2.1775\\
    $d^{\rm{geo}}, G$ & 1.2384  &  1.6523  &  1.7299  &  1.5613\\
    $d^{\rm{cost}},G$ & 1.2838  &  1.7246  &  1.8990  &  1.6471\\
    \hline
  \end{tabular}
  \label{TB:Radiation:Exponents}
\end{table}



\section{Discussion and conclusion}
\label{S1:Summary}

In this work, we investigated the highway freight transportation diversity of 338 Chinese cities based on the transportation probability $p_{ij}$ from one city to the other. The transportation probabilities are calculated from the raw radiation model based on geographic distance and the cost-based radiation model based on driving distance as the proxy of cost. 

We found that, in either the raw radiation model or the cost-based radiation model, the results obtained with the population and the gross domestic product are quantitatively similar. It is mainly due to the nice power-law scaling between population and GDP of Chinese cities, where the power-law scaling exponent is estimated to be $1.15\pm0.08$ \cite{Bettencourt-Lobo-Helbing-Kuhnert-West-2007-PNAS,Wang-Ma-Jiang-Yan-Zhou-2019-EPJDataSci}.

We investigated several important properties of the truck transportation probability $p_{ij}$. It is found that the transportation probabilities are distributed broadly with a nice power-law tail and the tail exponents are close to 0.5 for the four models. It is also found that the transportation probability matrix in each model is asymmetric such that $p_{ij}$ does not necessary equal to $p_{ji}$, which is consistent with our intuition.

We also found that the population, the gross domestic product, the in-flux, and the out-flux scale as power laws with respect to the transportation diversity in the raw radiation model and the cost-based radiation model. It is intuitive that a city with higher GDP (often with larger population) usually has higher diversity in its industrial structure. These cities usually have higher diversity in highway freight transportation.

\section*{Acknowledgements}

  This work was partly supported by the Fundamental Research Funds for the Central Universities.


%

\end{document}